# The Future of Stardust Science

A. J. Westphal[1], J. C. Bridges[2], D. E. Brownlee[3], A. L. Butterworth[1], B. T. De Gregorio[4], G. Dominguez[5], G. J. Flynn[6], Z. Gainsforth[1], H. A. Ishii[7], D. Joswiak[3], L. R. Nittler[8], R. C. Ogliore[9], R. Palma[10,13], R. O. Pepin[10], T. Stephan[11], M. E. Zolensky[12]

[1]Space Sciences Laboratory, U. C. Berkeley, Berkeley CA, westphal@ssl.berkeley.edu
[2]Space Research Centre, Dept. of Physics & Astronomy, University of Leicester, UK
[3]Department of Astronomy, University of Washington, Seattle WA
[4]U. S. Naval Research Laboratory, Washington DC
[5]California State University at San Marcos, San Marcos, CA
[6]State University of New York, Plattsburgh, NY
[7]Hawaii Institute of Geophysics and Planetology, University of Hawaii at Manoa, Honolulu, HI
[8]Department of Terrestrial Magnetism, Carnegie Institute of Washington
[9]Washington University in St. Louis, St. Louis MO
[10]Department of Physics, University of Minnesota, Minneapolis, MN
[11]Department of the Geophysical Sciences, The University of Chicago, Chicago, IL
[12]ARES, NASA JSC, Houston, TX
[13]Department of Physics and Astronomy, Minnesota State University, Mankato, MN

**Abstract**
Recent observations indicate that >99% of the small bodies in the Solar System reside in its outer reaches – in the Kuiper Belt and Oort Cloud.  Kuiper Belt bodies are probably the best preserved representatives of the icy planetesimals that dominated the bulk of the solid mass in the early Solar System.  They likely contain preserved materials inherited from the protosolar cloud, held in cryogenic storage since the formation of the Solar System.  Despite their importance, they are relatively underrepresented in our extraterrestrial sample collections by many orders of magnitude (~$10^{13}$ by mass) as compared with the asteroids, represented by meteorites, which are composed of materials that have generally been strongly altered by thermal and aqueous processes.  We have only begun to scratch the surface in understanding Kuiper Belt objects, but it is already clear that the very limited samples of them that we have in our laboratories hold the promise of dramatically expanding our understanding of the formation of the Solar System.   Stardust returned the first samples from a known small solar-system body, the Jupiter-family comet 81P/Wild 2, and, in a separate collector, the first solid samples from the local interstellar medium.   The first decade of Stardust research resulted in more than 142 peer-reviewed publications, including 15 papers in *Science.*  Analyses of these amazing samples continue to yield unexpected discoveries and to raise new questions about the history of the early Solar System.  We identify 9 high-priority scientific objectives for future Stardust analyses that address important unsolved problems in planetary science.

# INTRODUCTION

2016 marks the 10th anniversary of the return to Earth of the Sample Return Capsule from Stardust, the first mission to capture and return solid extraterrestrial samples from beyond the Moon. Stardust returned the first *cometary solids* ever sampled directly from a comet's coma, and a separate collector that was exposed to the stream of interstellar dust and which probably (but not yet certainly) contains a handful of tiny *local interstellar rocks*, the first ever identified. This low-cost, high-risk mission – really two missions, in one spacecraft – has been extraordinarily successful by any measure. The first decade of Stardust research resulted in more than 142 peer-reviewed publications, including 15 papers in *Science.* Unexpected discoveries have led to major revisions of our understanding of structure and history of the early Solar System and analyses of these amazing samples continue to yield unexpected discoveries and to raise new questions about the history of the early Solar System. For example, the discovery of highly refractory minerals, minerals implanted with high concentrations of implanted noble gases, and "microchondrules" implies large-scale outward transport of material in the disk, from the inner solar system all the way out to the Kuiper Belt. Like archaeologists investigating human origins through the study of rare ~4 Ma old fossils of ancient hominids in Ethiopia, cosmochemists are investigating our own cosmic origins through the study of rare ~4.6 Ga old "fossil" solar-system building materials, going a thousand times further back in time than our archaeologist colleagues. One Stardust investigation, still ongoing, inspired a multitude of "citizen science" projects that productively collaborate with thousands of amateurs in achieving important science goals across a variety of disciplines.

Because they are a rare, finite resource, Stardust samples pose logistical and technical challenges beyond those involved in analyzing most other extraterrestrial materials, such as meteorites, which are far larger and far more abundant. The analysis of a single complex 10 μm diameter particle, typically <1 ng ($10^{-9}$ g) in mass, can easily consume a year or more of effort by a small research group. Long-term investment by NASA and other agencies in improvements in analytical instruments and techniques, and in sample handling and preparation, which started with Apollo, has yielded spectacular advances in our understanding of the early Solar System, but it is equally clear that the research on Stardust samples, to date, has only scratched the surface and many fundamental questions remain about the nature of these samples. With Stardust cometary samples, we are now perhaps where we were in approximately 1960 in the study of meteorites. For example, we do not know the bulk composition of comets with sufficient accuracy to be able to make a sensible comparison with other extraterrestrial materials. A full understanding of the nature of these astonishing samples will require patience and perspective.

With dust particles from the Stardust interstellar collector, the situation is even more striking. Here we are perhaps where we were in approximately 1800 with meteorites, just after they were proposed to be cosmic in origin by Ernst Chladni, because it is not yet proven that any of the Stardust interstellar dust candidates identified so far are in fact interstellar in origin. Fundamentally, the reason is simple: as small and challenging as the cometary samples are, interstellar dust samples are several orders of magnitude more challenging. Typical interstellar dust particles are at least a thousand times less massive than typical cometary particles, and the total number of particles is <<1% of number of particles in the cometary collection.

A decade on, this is perhaps a good time to ask three questions: *What have we learned from the Stardust mission? What role will the analyses of Stardust samples play in the future of Planetary Science and Astrophysics? What resources will be needed to carry out these analyses?* We will organize the responses to the last two questions around a list of unsolved problems in planetary science and astrophysics that can be addressed by analyzing Stardust mission samples. The goal of this paper is to provide an overview of work that has and is being done on the comet and interstellar dust samples and provide some insight into investigations that can be done in the future. The focus is on Stardust mission samples but many of findings and discussions are clearly important to the analysis of nonvolatile components that will be obtained on future sample return missions.

This paper is the product of a workshop that was held near the UC Berkeley campus on July 26-27, 2015, just prior to the 78th annual meeting of the Meteoritical Society. The workshop was supported by NASA and sponsored by the Curation and Planning Team for Extraterrestrial Materials (CAPTEM), a NASA advisory committee.

**STARDUST SAMPLES IN CONTEXT**

A central goal of the field of cosmochemistry is to understand the formation and evolution of the Solar System based on clues preserved in ancient and primitive extraterrestrial materials. The period of formation of the major and minor planetary bodies during the first few million years of Solar System history is particularly challenging because this period can only be studied using clues derived from the analysis of relatively rare samples of material that formed at that time, and that have avoided subsequent alteration or annihilation due to exposure to heat or water. The clues are uncovered through use of sophisticated laboratory instruments that must be well-suited to the demanding nature of the samples. Beginning in the 1950's, and dramatically accelerated by investments during the Apollo era, the modern field of cosmochemistry cut its teeth on the study of meteorites, which are by many orders of magnitude the most readily available extraterrestrial samples. Thanks to ongoing commitment to meteorite collection campaigns such as ANSMET (Antarctic Search for Meteorites, active since 1976), many tons of meteorites exist in the worldwide collections (although to be sure some rare or unique meteorites exist in only ~gram quantities). Mineralogic and petrographic studies, measurements of elemental and isotopic composition, analyses of organics and noble gases, and others, have now reached an extraordinary level of sophistication. One consequence of these improvements was a dramatic reduction in the required size of samples for a given type of measurement, sometimes by orders of magnitude. Improvements in sensitivity and resolution led to discoveries that otherwise would have been impossible, such as the discovery of trace quantities of ancient "presolar" grains in primitive meteorites (e.g., Zinner 2014).

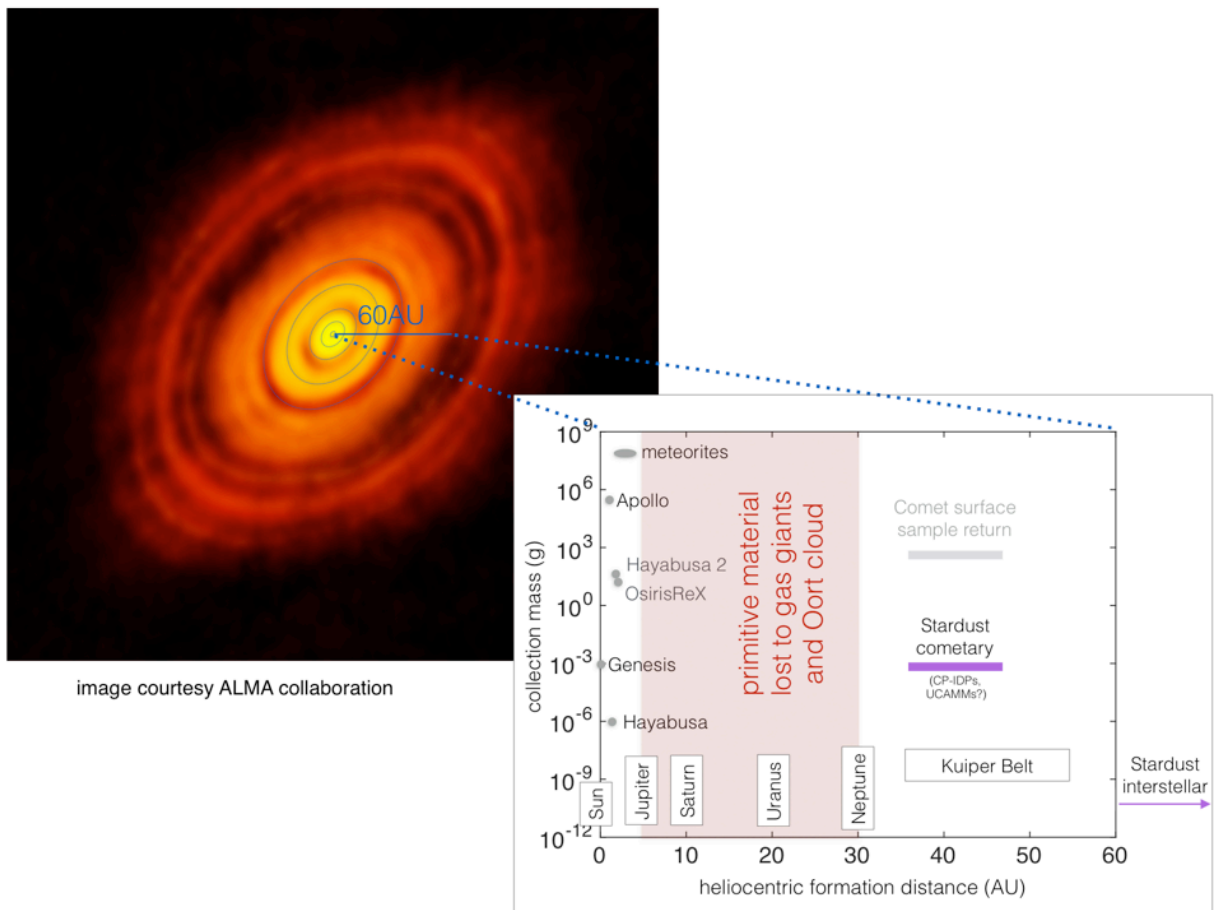

Fig. 1. Masses of extraterrestrial sample collections versus heliocentric distance of formation. Existing collections are shown in bold, collections from missions in progress (Hayabusa 2, OSIRIS-REx) or proposed (New Frontiers Comet Surface Sample Return) are shown in grey. The mass of the Stardust cometary sample is tiny compared to the meteorite collection but is the first known to have originated outside the inner Solar System. Extraterrestrial sample collections suffer from a severe sampling bias: since Kuiper-belt objects outnumber asteroids by a factor of ~100, they are relatively undersampled as compared with the asteroids by $>10^{13}$. For a sense of potential scale, the recent millimeter-wave ALMA image of the protoplanetary disk HL Tau is shown, overlaid with the orbits of the Earth and the four solar-system gas giants. It is not yet known whether the Solar System's protoplanetary disk was as extensive as this one. HL Tau image credit: ALMA (ESO/NAOJ/NRAO).

Meteorites are samples of small bodies that are thought to have formed in a relatively narrow range of heliocentric distances, mostly between 2 and 3 AU, between the present-day orbits of Mars and Jupiter. The remarkable mineralogical, chemical, and isotopic complexity of the meteorite collection is itself an important observation, and implies strongly varying conditions in space (and probably time) in the inner Solar System, with admixing of equally complex material formed elsewhere. However, astronomical observations of protoplanetary disks, and especially spectacular new millimeter-wavelength observations by ALMA (Figure 1, Atacama Large Millimeter/submillimeter Array, almaobservatory.org), show protoplanetary disks extending to 100 AU or more.  The extent of the protoplanetary disk from which our Solar System evolved is unknown, but it is likely that meteorites sample only a tiny, unrepresentative fraction of the original protoplanetary disk.  Asteroids themselves are now thought to constitute only a small fraction of the population of small bodies in the Solar System; recent observations indicate that Kuiper Belt objects, which formed or at least orbit beyond Neptune, outnumber asteroids by at least two orders of magnitude (Schlichting et al. 2012, Bottke et al. 2005, Farinella and Davis 1996).  Analysis of samples formed in outer regions of the disk is thus required for even a partial understanding of the structure and history of the Solar System.

We now know of several thousand extrasolar planets. It is clear from the data that there is a wide variation in the types of planets and distribution of these within extra-solar planetary systems. How this variation came to be is a very important area of astrophysics and is a question that analyses of the oldest and arguably most pristine samples of Solar System building blocks can help to answer.

Starting in 1981, immediately after the pioneering work of Don Brownlee (Brownlee et al., 1977), the then-future PI of the Stardust mission, NASA has been collecting Interplanetary Dust Particles (IDPs) in the stratosphere (Warren and Zolensky, 1994).  Several lines of evidence point to a cometary origin for a class of fine-grained IDPs containing anhydrous minerals, called Chondritic-Porous IDPs (Bradley & Brownlee 1986, Bradley 1994, Bradley *et al.* 1999). Because of limitations of the collection technique, no particular IDP can be unambiguously identified with a specific small body, so the question of their origin(s) is not yet completely resolved.  Larger particles called micrometeorites (e.g., Dobrica *et al.* 2009, Noguchi *et al.* 2015) have been recovered from polar ice and snow (Maurette 1991, 1994, Duprat *et al*. 2010).   These larger particles are generally more strongly heated during atmospheric entry and may be altered in the terrestrial environment, especially by leaching.  As a consequence, their origins are less certain, although they probably derive from both asteroids and comets.  Some have been found to have identical fine-grained components to Chondritic-Porous IDPs and, thus, sample sources that are most likely cometary (Noguchi et al. 2015).

The Stardust mission collected materials from the coma of the Jupiter-Family comet Wild 2 in 2004 (Brownlee *et al.* 2003a).  The total mass of the collection, >300 µg (Hörz *et al.* 2006), is orders of magnitude smaller than even the smallest meteorite.  But despite its diminutive size, the "lever arm" of this sample collection is enormous.  Although the Stardust spacecraft collected material from Wild 2 while it was in the inner Solar System, Wild 2 almost certainly formed in the Kuiper Belt (it has been in its present orbit only since 1975, when it had a close encounter with Jupiter). So the Stardust cometary samples are the first from a known solar-system body that likely formed in the outer Solar System (Fig. 1).  Recent studies of gas emitted from Wild 2 show that it has a frozen volatile content that is typical of comets from the Kuiper Belt (Dello Russo et al. 2014).  Because we have been studying meteorites for decades and have literally tons of material in our collections, we have a bias towards thinking of these as typical of small solar-system objects, though in fact, they are not. More than 99% of small bodies in the Solar System probably resemble objects like Wild 2: recent observations imply that Kuiper Belt Objects outnumber asteroids by more than two orders of magnitude both in number and by mass (Schlichting *et al.* 2012, Bottke *et al.* 2005).

|  | planets and differentiated asteroids | undifferentiated asteroids | comets | interstellar dust |
|---|---|---|---|---|
| **Parent body** |  |  |  |  |
| melting and differentiation | major | minor to none | none |  |
| parent-body thermal metamorphism | major | major to minor | none |  |
| aqueous alteration | major | major to minor | moderate to none |  |
| shock | major to minor | major to minor | minor to none |  |
| **Nebula** |  |  |  |  |
| hot evaporation and condensation |  | major | minor? |  |
| chondrule-forming process or other nebular heating |  | major | major? |  |
| radiation processing |  | minor? | major? (e.g., GEMS) |  |
| cold condensation and evaporation |  | minor? | major? (e.g., organics) |  |
| **ISM** |  |  |  |  |
| hot evaporation and condensation |  |  | minor? (e.g., enstatite whiskers) | major? |
| radiation processing |  |  | major? (e.g., GEMS?) | major? |
| cold evaporation and condensation |  | minor? | major? (e.g., GEMS?) | major? |
| **Circumstellar** |  |  |  |  |
| hot evaporation and condensation |  | minor | minor? (e.g., presolar grains) | minor? |
| radiation processing |  | minor? | minor? | minor? |
| cold evaporation and condensation |  | minor? | minor? | minor? |

Table 1.  Processes that control chemistry and mineralogy of extraterrestrial solids, , and which may be recorded and preserved in them.

Interest in the degree to which our extraterrestrial sample collections are representative of all Solar System bodies is driven by the different environments they likely sampled and the different processing they underwent. In Table 1, we highlight the different processes affecting classes of planetary bodies and interstellar materials.   The processes that control the mineralogy of cometary and interstellar dust overlap those that control the mineralogy of meteorites, but generally those that are minor in asteroids are major in comets, and vice versa.

**Advances in laboratory instrumentation**

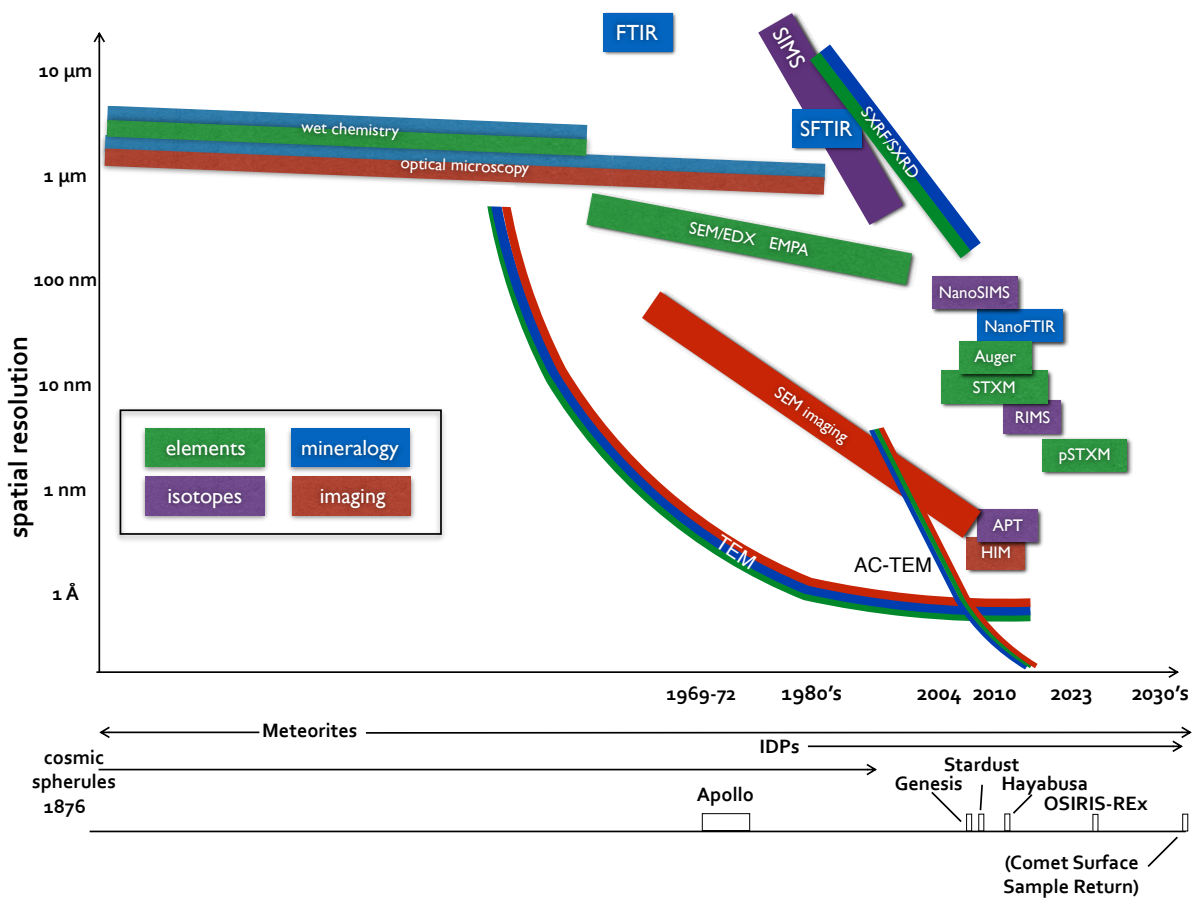

Fig. 2  Illustration of the advances in spatial resolution of various microanalytical techniques, superimposed on a timeline of returned extraterrestrial samples.   See Appendix for glossary of terms.

Unlike most NASA missions, the instrument suite of the Stardust mission was relatively simple. The mission carried particle impact detectors and a time-of-flight impact mass spectrometer but its primary goal was to use passive collectors to capture and return solid materials from the coma of a specific comet, 81P/Wild 2, and, in a separate collector, interstellar dust from the local interstellar dust stream. Laboratory identification of phases in Wild 2 samples is generally more straightforward and diagnostic than identification through remote observations in cometary comas. In this sense, Stardust provides "ground truth" for astronomical observations. Most of the analytical sophistication of the mission was – and continues to be – on the ground after recovery. Researchers analyze samples using instruments that could never be flown in space, such as synchrotrons – X-ray and infrared microprobes the size of shopping malls. Instruments on Earth are unconstrained by the mass, power, and reliability requirements imposed by spaceflight and need not exist at launch. With careful curation (Zolensky et al. 2008a, Frank et al. 2013), samples can be analyzed decades after return, taking advantage of instruments with capabilities that might not be imagined during mission planning. Perhaps most importantly, measurements can be repeated by different instruments in different laboratories to confirm (or disprove) results – the backbone of the scientific process. Results from analyses often lead to refined approaches to investigate issues that were not previously known. The discovery of presolar grains in meteorites is an excellent example of what can ultimately evolve from complex, often confusing, and ever improving data.

**<u>Challenges in Stardust sample analysis</u>**
Stardust samples pose challenges that are unfamiliar to those accustomed to studying other sample collections. Unlike the meteorite, IDP, or polar micrometeorite collections, which are more readily available because new samples are continually raining from the sky, the samples returned by Stardust are a small and finite resource. As a consequence, the process for requesting, preparing, and allocating samples to researchers is deliberately painstaking and can be frustratingly slow. The highly demanding nature of the samples requires cutting-edge instruments that are often highly oversubscribed, as well as the development of novel sample handling techniques and procedures to prepare the samples for analysis in those instruments. While the Stardust cometary collection suffers from this challenge to a certain extent, the interstellar dust collection demonstrates it in the extreme. The total number of "large" particles, of order 1 picogram ($10^{-12}$g) in mass, is probably no more than a dozen, and the total mass of the interstellar collection is estimated to be less than one-millionth that of the Stardust cometary collection.

Although great sample-derived insight on the origin of comets has come from the first decade of study, it is clear that the research on Stardust samples has only scratched the surface. We do not yet understand the diversity of materials that compose even this one comet, Wild 2. It appears that Wild 2 contains abundant "xenoliths" (foreign rocks) from all over the Solar System, including minerals that form at high temperatures (Zolensky et al. 2006); however, observations of the stratified structure of comet Churyumov-Gerasimenko by the ESA Rosetta mission implies that its formation was an episodic, gentle, and cold process (Massironi et al.

2015).. Is transport in a turbulent disk consistent with both observations, and could these "xenoliths" really have been transported in such large quantities from the inner Solar System? Or is it possible that they were somehow formed in the outer Solar System?

Answers to these questions will yield important advances in our understanding of the early Solar System, similar to those resulting from the Apollo-era-initiated, long-term NASA investment in analytical instruments and in sample preparation. More than 95% of the samples returned in 2006 are still available for analysis.

The Stardust mission also served as a catalyst for the development (or refinement) of analytical techniques to maximize the scientific payout of the mission. Because there are other sample return missions, either in flight or in the proposal stages (OSIRIS-REX, Hayabusa 2, Comet Surface Sample Return), the techniques and expertise developed through NASA support of laboratory analysis of returned samples will continue to be essential to the success of these missions. Investment into instrumentation capable of processing cometary samples yields a double dividend as the same instrumentation advances are applied to meteorites and IDPs.

**SCIENTIFIC OBJECTIVES FOR THE NEXT DECADE OF STARDUST RESEARCH**

Here we summarize our view of the highest priority scientific objectives in the coming decade for analyses of Stardust samples. Although a great deal was learned from the remote observations of comet Wild 2 during the flyby by the Stardust spacecraft, we focus here on the primary goal of the mission: discoveries arising from the first laboratory analyses of *bona fide* cometary solid materials collected from a known comet whose activity and orbit is typical of Jupiter family comets. Because the interstellar collection is several orders of magnitude more challenging than the cometary collection, even after a decade we are still in the infancy of its analysis, and there are many uncertainties;the evidence for interstellar particles in the aerogel collector is still circumstantial and remains to be confirmed through further analyses.

**Objective 1:** *Determine modal mineralogy of Wild 2 for comparison with primitive chondrites and CP-IDPs*

| Question and Objective | measurement | instrument | sample requirements | sample frequency |
|---|---|---|---|---|
| **1 What are comets made of?** *Determine modal mineralogy for comparison with primitive chondrites and CP-IDPs* | Major modal mineralogy (coarse grains, fines, Type II chondrules and chondrule fragments, sulfides, metal, Kool grains, organics, carbonaceous material), bulk chemistry | analytical TEM, µXRF/XRD, XANES, EBSD | keystone, ultramicrotome sections | common |
| | Concentration of minor phases (CAIs, Type I chondrules, Al-rich chondrules, enstatite whiskers, cubanite, pentlandite) | analytical TEM, µXRF/XRD, XANES | ultramicrotome sections | common |
| | Concentration of phases exclusive to CP-IDPs (enstatite whiskers, GEMS, Kool grains) | analytical TEM | shielded material (possibly behind terminal particles) | rare (Kool grains) unknown (GEMS, whiskers) |
| | Concentration of primary condensation products (LICE/LIME olivine, enstatite whiskers) | analytical TEM, µXRF/XRD | ultramicrotome sections | common |

**Summary of objectives**

- **Determine the relationship of Wild 2 CAIs and meteoritic CAIs**
- **Determine the relationship between Wild 2 microchondrules and chondrule fragments and meteoritic chondrules**
- **Clarify the relationship between Wild 2 and CP-IDPs**
- **Accurately determine the modal abundance of CAIs**
- **Accurately determine the relative mix of coarse-grained and fine-grained components**
- **Substantially improve statistics of analyses of Kool grains**
- **Accurately determine the presolar grain abundance**
- **Accurately determine the enstatite whisker abundance**
- **Accurately determine the LIME forsterite abundance**
- **Accurately determine the GEMS and Polyphase Unit abundance**
- **Determine the abundance of aqueous alternation products (Objective 5)**

**Technology advances**
- **High energy-resolution X-ray spectroscopy (e.g., microcalorimetry)**
- **High spatial-resolution synchrotron mapping with Mg, Al Kα detection**
- **Improve dust particle accelerator technology to reduce acceleration shock**

When taking a first look at the Wild 2 samples a decade ago, it was natural to ask: *Have we seen this material before?* Can we recognize materials that we find in other primitive solar-system materials – in primitive meteorites or in primitive CP-IDPs?

The process of answering this straightforward question is complicated by several confounding issues. All three collections – primitive meteorites, CP-IDPs and Wild 2 cometary samples – have been altered since their formation ~4.6 Gy ago by distinct mechanisms and to varying degrees. Even the most primitive meteorites show signs of aqueous alteration or thermal metamorphism that took place on the small bodies from which they originated; the finest-grained components (the "matrix") in meteorites is the most susceptible to these effects. CP-IDPs were often heated to incandescence during atmospheric entry, so are thermally altered and oxidized to varying degrees by reaction with atmospheric oxygen, particularly on their rims. Wild 2 samples are altered in complicated ways by the effects of capture at 6.1 km/sec in aerogel or aluminum foil. Nevertheless, an inventory of the materials in returned Wild 2 samples that allows for a first-order comparison to the other collections sets the context for understanding them.

**Major modal mineralogy**

The early work on the Wild 2 samples involved a learning curve to develop means of distinguishing between well-preserved materials and those that had been altered during high speed capture (Brownlee *et al.* 2006). Although laboratory capture experiments had been done previously, this was the first experience with cometary solids, a complex fragile mix of fine and coarse components. Generally, it was found that components near the ends of aerogel capture tracks and those larger than a few micrometers were well preserved, often with a protective cap of compressed $SiO_2$ aerogel. However, all tracks also contained components that were modified to various degrees from slight to severe in which some components melted and dissolved in the molten silica that lines the walls of capture tracks (Ishii *et al.* 2008a).

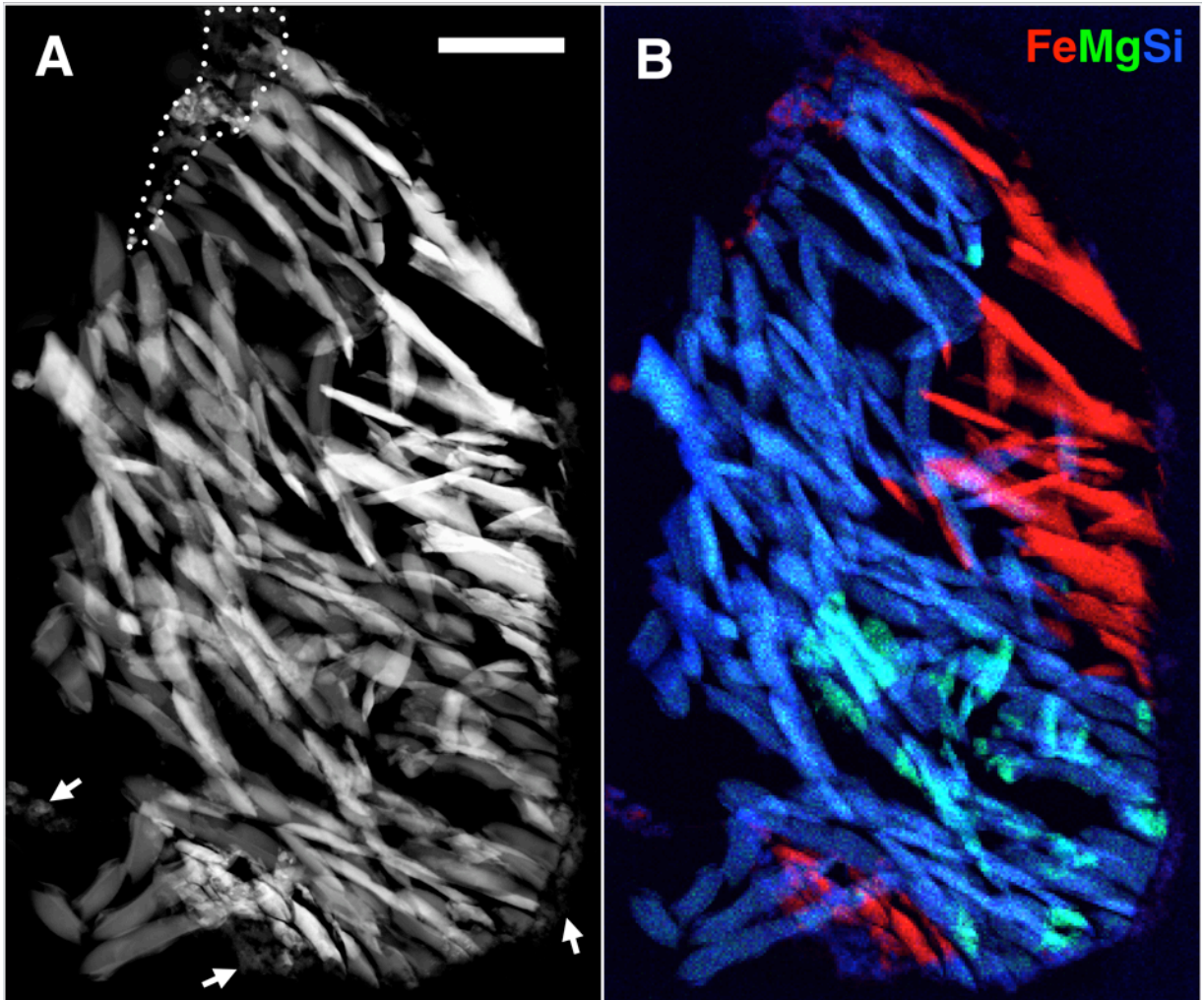

Fig. 3: TEM view of a Stardust terminal particle ultramicrotomed section. A) STEM HAADF image. A large region of mixed aerogel and Stardust is outlined at the top, and additional "fusion crust" regions are marked with arrows. 1 micron scale bar. B) EDS view of the same region with Fe (red), Mg (green), Si (blue) showing sulfide in the upper corner and bottom, and silicate everywhere else.

One of the very first findings of the mission was that the comet was composed of a wide range of particles that ranged from solid grains, which produced long thin tracks (Type A) made by non-fragmenting material, to loose fines or unstable materials, which produced broad (bulbous) tracks (Trigo-Rodriguez *et al.*, 2008, Burchell *et al.*, 2008a). Some of the solid grains were larger than 50 µm with beautifully preserved interiors (e.g., Nakamura et al. 2008).. One interpretation of the capture process is that the thermal capture effects only influenced the outer micrometer or less and that larger grains were much better preserved than smaller ones (Brownlee, Joswiak and Matrajt 2012). The situation is somewhat analogous to meteorites, where entry heating melts a thin fusion crust but thermal inertia protects their interiors.

Among early observations was the remarkable finding that isotopically anomalous ("pre-solar") grains are rare and that the comet contains fragments of common high temperature nebular materials, typically found in meteorites.  The discovery of micro-chondrules and micro-CAIs (Ca-Al-rich inclusions) in the comet was quite unexpected and totally at odds with the common belief that comets formed in isolation from the inner Solar System (McKeegan et al. 2006, Simon et al. 2008, Chi et al. 2009).  In initial studies, the terms "chondrule-like" and "CAI-like" were used because it was not clear that the usually <50 µm fragments were indeed fragments of the same components found in meteorites.  After nearly a decade of effort, the similarities are better established: so far all chondrule-like fragments more closely resemble type II chondrules than Type I chondrules (e.g., Gainsforth et al. 2015),  but it remains to be determined if the Wild 2 chondrules and refractory inclusions do actually sample the same populations of these nebular components found in chondrites.

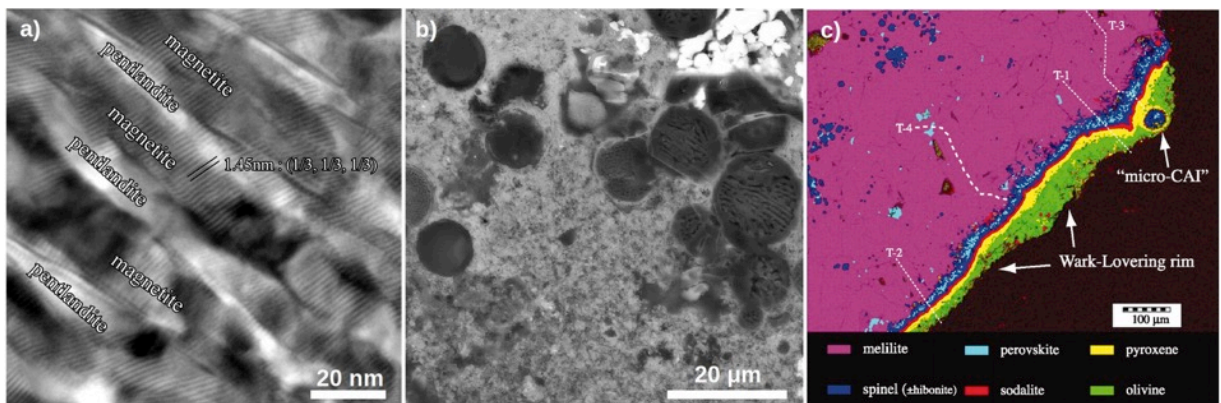

Fig. 4:  a) HRTEM images of cosmic symplectite (composed of "wormy" grains of magnetite and pentlandite) in Acfer 094 (ungrouped carbonaceous chondrite), from Seto et al., GCA, 2008.  b) Backscattered electron image of micro-chondrules in the unequilibrated ordinary chondrite MET 00526 (image credit: Elena Dobrica, UNM).  c) Compositional map (from EPMA data) of Wark-Lovering rim with an included micro-CAI, from Simon et al., Science, 2011.

Through a *tour de force* of mission design, the closing speed of the Stardust spacecraft with comet Wild 2 was only 6.1 km/sec.  This relatively low speed was extremely important for sample science.   Because it is within the speed range of two-stage light-gas guns, experiments can be readily carried out to assess the effect of capture in aerogel (Burchell *et al.* 2009, Burchell *et al.* 2008a, Burchell *et al.* 2012, Kearsley 2012) and aluminum foils (Kearsley 2006, Burchell *et al.* 2008b, Kearsley 2008, Wozniakiewicz *et al.* 2009, Kearsley 2009, Price 2010, Wozniakiewicz *et al.* 2011, Wozniakiewicz *et al.* 2012a, Wozniakiewicz *et al.* 2012b, Wozniakiewicz *et al.* 2015, Foster  *et al.* 2013, Hörz 2012, Nixon 2012)  at the Wild 2 encounter speed.  These experiments, in coordination with numerical simulations, allow for determination of the speed, trajectory and compositional distribution of ejecta on various materials, which has been important in understanding the origin and distribution of secondary particles captured in the interstellar collector of Stardust (Burchell *et al.* 2012, Price 2012).  However, because the acceleration process itself subjects samples to strong shocks

(Bogdanoff, private communication), it is not always possible to unambiguously distinguish capture effects from acceleration effects (Stodolna 2012b). There are other factors – notably mineral density and hardness – which are also likely to have influenced the preservation of different mineral species during the capture within aerogel, and these remain to be fully understood. **Improvements are needed in dust acceleration technology, particularly in developing the capability of accelerating fragile cometary particles more gently.**

A major fraction of the coarse grains (that is, those >2µm in size) may be fragments of chondrules that bear some chemical and isotopic similarities to those found in chondrites (Brownlee *et al.* 2012). Wherever and whenever they formed, it is clear that they formed by similar high temperature processes that formed chondrules and that they have dramatically different origins than previously expected for cometary grains (Gainsforth *et al* 2015). Although accurate abundance values are difficult to determine in Stardust samples due to the loss of petrographic context and complexities of fragmentation, the abundance of micro-CAIs (by number) is on the order of 2% and the abundance of chondrule-like fragments is much higher (Joswiak et al. 2012). The chondrules and CAIs in Wild 2 provide direct evidence for profound large scale mixing in the early Solar System, unless there was a high-temperature process in the outer Solar System that was a source of materials compositionally, isotopically, and mineralogically similar to those made in the inner Solar System.

*Fines*

An overriding issue with the Wild 2 samples is the nature of the components that either poorly survived capture or did not survive at all. This includes micrometer-sized and smaller components that often stopped in the upper regions of tracks and less robust materials that were not amenable to capture. Although many Wild 2 submicrometer-sized grains have been studied, the nature of typical fines remain poorly determined or unknown, mainly due to the possible severe effects of alteration during the capture in aerogel or on Al foils. While many of the best studied tracks contain an abundance of well-preserved coarse materials, the contents of many of the very largest (bulbous) tracks remain enigmatic. Analyses have generally focused on the well preserved larger components, because they are more amenable to study, and addressing high-level science questions has been hampered by both the human and capture selection biases disfavoring fines. For example, isotopically anomalous presolar grains were expected to be in high abundance in comets prior to the return of the Stardust samples, but were in fact to be found to be quite rare (Stadermann et al. 2008). Extensive experimental efforts subsequently indicated that hypervelocity capture on Al foils often destroys presolar grains leading to anomalously low measured abundances (Floss et al. 2013, Croat et al. MAPS 2015). This study has not yet been done for particles captured in aerogel. Similarly, it is not known how the isotopic and chemical properties of small organic-rich grains collected along tracks have been affected by the capture process (see Objective 4). The uncertain nature of the fines affects our understanding of the bulk elemental composition of the comet solids as well as the presence of critically important materials, discussed below, that may provide direct genetic links to other meteoritic samples.

**The relative concentrations of coarse robust and fine or fragile components is an important measurement that has only been addressed by existing Wild 2 studies in a preliminary way.** The comet is a container of collected nebular materials that were probably transported to the Jupiter family comet accretion region, although the question of *in situ* formation remains open. The fines are an important component that has been difficult to study and have only been directly studied by a few investigators(e.g., Ogliore et al. 2015). This "missing component" should be emphasized in future work so that we can better understand the extent of nebular and Solar System components in comets.

## Concentrations of minor phases

*Micro-CAIs*

A multiyear TEM study of 19 Stardust tracks resulted in the discovery of highly refractory grains in at least five separate tracks. Mineral assemblages, mineral chemistries and measured bulk particle compositions (amongst other properties) demonstrated that these grains are similar to refractory materials in chondrites including fine-grained inclusions, (Krot *et al.*, 2004; Lin and Kimura, 1998; Simon et al. 2008; Chi et al. 2009; Joswiak et al. 2012) and Al-rich chondrules (Bridges *et al.*, 2012; Joswiak *et al.*, 2014). Except for the terminal particle in carrot-shaped track 130, a probable Al-rich chondrule fragment, the refractory grains were present in Type B/C bulbous tracks in which numerous other rock and mineral fragments were found, typically including olivines, pyroxenes, sulfides, oxides, and the ubiquitous fine-grained glassy material formed by impact melting. **A major goal is to determine the relationship of Wild 2 CAIs and meteoritic CAIs, and to determine accurately the modal abundance of CAIs.** (See also Objective 6)

*"Kool" grains*

A significant number of Stardust tracks contain assemblages consisting of FeO-rich olivines and Na- and Cr-rich clinopyroxenes (typically augites), sometimes with poorly crystallized albite or albitic glass with spinel (Joswiak *et al.* 2009). These assemblages have been named "Kool" (Kosmochloric high-Ca pyroxene and FeO-rich olivine) grains and are observed in more than half of all Stardust tracks including all nine bulbous tracks studied by TEM by the University of Washington group including track 25 (Inti), which contains > 20 CAI fragments. Kool grains are also observed in chondritic porous IDPs and are relatively common in a giant cluster interplanetary dust particle believed to have a cometary origin. The textures and mineral assemblages of Kool grains are suggestive of formation at relatively high temperatures and indicative of either igneous or metamorphic processes and may have formed under relatively high $f_{O2}$ conditions. Kool grains appear to be unique to comet Wild 2 samples and CP-IDPs; comparable mineral assemblages and mineral compositions have not been observed in chondrites. However, FeO-rich olivine andclinopyroxene assemblages in R chondrites, an oxidized chondrite class, somewhat resemble Kool grains, but have lower Na and Cr contents than the Kool grains (Joswiak et al. 2009).

The O isotopic composition of a single Kool grain from bulbous track 77 is comparable to some type II (FeO-rich) chondrule olivines from OC, R, and CR chondrites (Kita *et al.* 2010, Isa *et al.* 2011, Krot *et al.* 2006, Connolly and Huss 2010). (Tracks are numbered sequentially in the order in which they are extracted from the aerogel tiles.) No other isotopic measurements of Kool grains have been made. Because Kool grains are composed of moderately high temperature minerals, have fine grain sizes, possess approximately chondritic bulk major element compositions, and are common in cometary materials, these grains may represent an important contribution to precursor-type materials that formed objects such as FeO-rich chondrules. One type II microchondrule in Wild 2 shows kosmochloric enhancement (coupled Na and Cr substitution) possibly reinforcing the link between Kool grains and chondrule forming processes (Gainsforth 2015). Kool grains would not be expected to represent the earliest precursor types, however, and likely formed from pre-existing solids, possibly of nebular origin. **Further identification and analysis of Kool grains by TEM is a high priority, to understand their genesis.** The apparent absence of Kool grains in chondrites may indicate their destruction by parent body processes such as aqueous alteration.

*Micro-chondrules / chondrule fragments*

Several tiny igneous rocks from comet Wild 2 have been discovered and analyzed (Matzel et al. 2010; Joswiak et al., 2012, Ogliore *et al.* 2012b., Gainsforth *et al.* 2015, Nakamura *et al.* 2008a). Morphologically and texturally they are similar to chondrules found in meteorites, though significantly smaller (Fig. 5). There are also significant differences in the chemistry and mineralogy of Wild 2 micro-chondrules and meteoritic chondrules: the most common type of meteoritic chondrule is FeO- and volatile-poor chondrules, called Type I, whereas few micro-chondrules with such chemistry have been identified in Wild 2 (Joswiak *et al.*, 2012), although the comet does contain many Fe poor olivines that could be fragments of type I chondrules. Several examples of FeO-rich, volatile-rich micro-chondrules have been identified in Wild 2, while meteoritic chondrules with similar properties (of so-called Type II) are less common. An Al-rich, $^{16}$O rich chondrule fragment has also been identified, with similarities to those present in carbonaceous chondrites (Bridges et al. 2012). **Determining the relationship, if any, between microchondrules found in Wild 2 and chondrules found in chondrites is a high priority. Do they share a common origin, or are they only similar because they were once molten?**

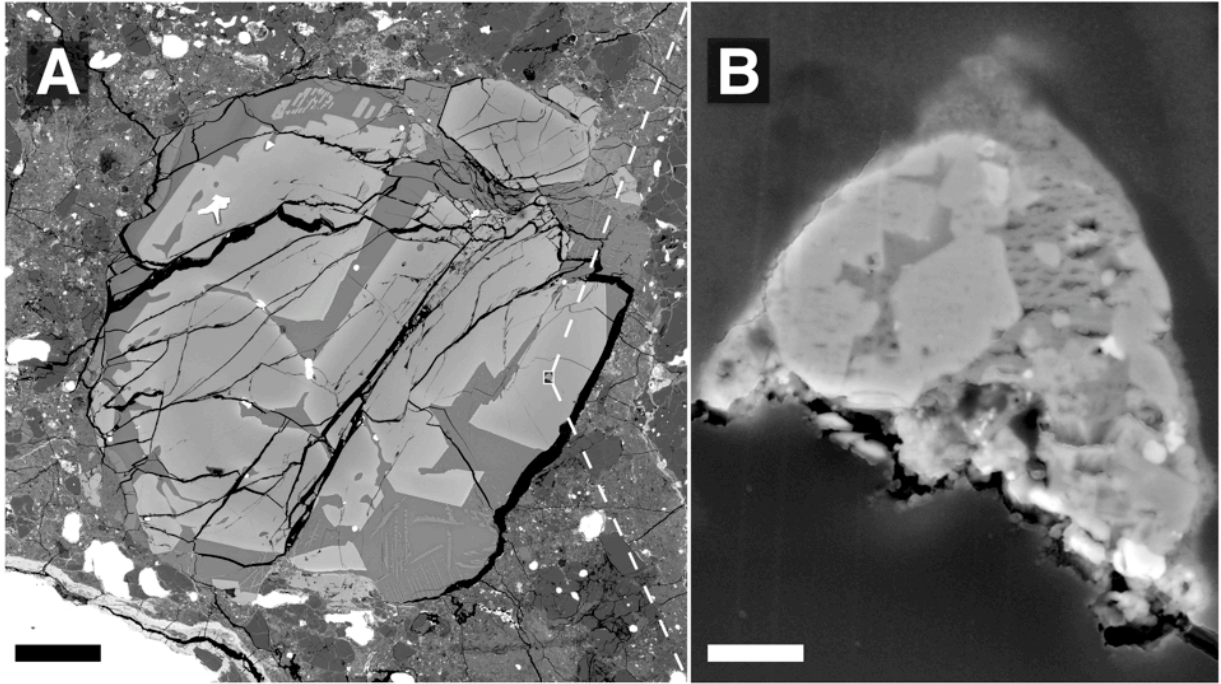

Fig. 5 Comparison of a typical chondrule with a Stardust "microchondrule". A) SEM BSE image of a type II chondrule from the CR2 meteorite GRA 95229, 100 micron scalebar. B) SEM BSE image of a terminal particle from Stardust named "Cecil," 2 micron scalebar.

**Components found in chondritic porous interplanetary dust particles (CP-IDPs)**

An important question is whether the samples returned from Wild 2 are typical of Kuiper Belt objects. Chondritic porous interplanetary dust particles (CP-IDPs) are commonly thought to derive from comets, so these serve as a touchstone for comparison with Wild 2. Properties that suggest CP-IDPs are cometary include their highly fragile porous aggregate structures, unequilibrated mineral assemblages, high carbon abundance, uncorrelated olivine Fe/Mn ratios, high presolar grain abundances as well as a number of other characteristics. Thus, CP-IDPs provide a unique opportunity to compare a likely independent comet sample to Wild 2.

While the comparisons are complicated by selection biases and capture effects, some statements can be made with confidence. Initial reports (e.g., Ishii et al. 2008) showed strong differences between Wild 2 samples and CP-IDPs; recent studies are showing affinities (e.g., Stodolna et al. 2014). **Clarifying the relationship between Wild 2 and CP-IDPs is a major goal of the next decade of Stardust research, because differences between them may give information about how and when comets were assembled and how representative Wild 2 is of other comets.** Nesvorny et al. (2010) showed that the distribution of zodiacal light is consistent with an origin of most interplanetary dust in whole-body break-up of Jupiter-family comets (JFCs); if this is correct, CP-IDPs sample the entire volumes of JFCs. The

Stardust cometary collector in contrast sampled material from the surface of JFC Wild 2. If they are different, and if Wild 2 is a typical JFC, then systematic differences may be indicative of radial heterogeneity in comets, preserving a stratigraphic record of the materials from which they were assembled, or perhaps reflecting processes altering the surface of comets at least to some thermal skin depth. The recent observation of apparent depositional layering in JFC comet 67P/Cheryumov-Gerasimenko (Massironi et al. 2015) may be consistent with this picture.

The *presolar grain abundance* in Wild 2 is challenging to measure because of strong capture effects in the Al foils, but after correction for capture effects, the concentration (~1000 ppm) appears to be in the middle or upper end of the range observed in CP-IDPs (Floss et al. 2013) and higher than in chondrites. However, both statistical and systematic uncertainties are large, so improvement in the accuracy of this measurement is a high priority for future work.

*Noble gases* have been identified in both IDPs and in Wild 2 samples. We discuss the question of the origin of noble gases under Objective 8.

*Enstatite whiskers*, which are thought to be condensation products, are rare components that are found in CP-IDPs, and were thought to be absent in Wild 2 samples (Ishii et al. 2008a). An elongated 300-nm long enstatite particle has been observed in Wild 2 (Stodolna et al. 2014), but the Wild 2 enstatite whisker was apparently embedded in a (non-aerogel) amorphous silicate matrix, unlike enstatite whiskers in CP-IDPs that are unembedded. It is possible that En whiskers are as abundant in Wild 2 as they are in andydrous IDP.

*LIME forsterites* (low-iron, manganese-enriched forsterites) with enhanced Mn/Fe ratios (>> 0.5%) are also thought to be condensates. They are commonly found in IDPs and in fact were first found in IDPs but also exist in chondrites, sometimes as components of amoeboid olivine aggregates (AOAs). The LIME olivines that have been found in Wild 2 and that have been analyzed for oxygen isotopic composition have primitive $^{16}$O-rich compositions similar to CAIs , AOAs and the Sun.

*GEMS (Glass with Embedded Metal and Sulfides)* is an enigmatic sub-micron phase ubiquitous in CP-IDPs. GEMS are commonly found in anhydrous chondritic-porous interplanetary dust particles and in some primitive and well-preserved micrometeorites. They are rarely, if ever, observed in meteorites and are believed to be a cometary component (Bradley 1994, Bradley *et. al* 1999, Zolensky et al. 2003). GEMS are easily destroyed by heating or aqueous alteration. A handful of objects have been discovered in the Stardust cometary collectors that are generally consistent in composition and morphology with GEMS in CP-IDPs (e.g., Gainsforth *et al.* 2016), but the capture processes also produced glass that mimics GEMS morphology. The hope is that indigenous cometary GEMS can be identified in the Stardust samples, possibly shielded from impact alteration by a robust terminal particle. To date, these GEMS-like objects have been found located along aerogel impact tracks and near terminal particles comprised, in part, by sulfides. This presents a major barrier to ruling out capture effects as the source of such objects: Stardust speed impacts result in ablation, abrasion, melting and intimate co-mingling of aerogel and mineral impactors, and those that involve sulfides generate the nanometer-scale sulfide and metal beads characteristic of CP IDP GEMS (Ishii *et al.* 2008a, Abreu *et al.* 2011, Ishii & Bradley 2015). Thus, it has not yet been definitively proven that Wild 2 contains GEMS.  This is a critical issue because one of the suggested origins for GEMS is that they are presolar interstellar grains that generally do not have isotopically anomalous compositions because of homogenization processes that occur in the interstellar medium (Bradley 1994, Bradley & Ishii 2008, Villalon *et. al* 2016).  It is possible that Wild 2 contains an abundance of interstellar isotopically normal GEMS that are unrecognizable because they were small and dissolved into molten silica when they came in contact with aerogel during capture.

*Coarse grains* A study of ~50 coarse grains from a single giant CP-IDP consisting of hundreds of coarse grains ~5 – 40 μm, revealed the presence of four refractory particles (Joswiak and Brownlee, 2015).  Similar to those in Wild 2, the grains consist of moderately refractory materials and include 3 fine-grained CAIs and a probable amoeboid olivine aggregate. Like Wild 2, the highest temperature refractory inclusions (e.g., hibonite) in the giant cluster particle appear to be absent.   Two of the refractory inclusions in the giant cluster IDP consist of individual concentric nodules composed of anorthite/spinel interiors with Al,Ti clinopyroxene rims with the nodules cemented together in a texture similar to  fine-grained refractory inclusions in chondrites and somewhat akin to Inti, the first CAI found in Wild 2 samples.  The discovery of CAIs in Wild 2 also represents one of the most useful results of the Stardust Mission: it will permit us to re-examine the existing IDPs and micrometeorites that we have collected for 40 years and better understand which might originate from comets. When refractory IDPs were first recognized, 30 years ago (Zolensky, 1987, McKeegan, 1987) they were assumed to be of asteroidal origin. Now we can look at these samples with new eyes, leveraging thousands of other samples collected by NASA.

**Objective 2: Test the extent of large scale nebular transport and mixing**

| Question and Objective | measurement | instrument | sample requirements | sample frequency |
|---|---|---|---|---|
| 2 How does material move and evolve in protostellar disks? *Determine extent of large-scale transport and mixing* | Concentration of primary condensation products (LICE/LIME olivine, enstatite whiskers) | analytical TEM, µXRF/XRD | | |
| | Complementarity between fines and large components | µXRF | keystone | common |
| | Distribution of olivine Fe content in microchondrules | analytical TEM | u/m sections | common |
| | Crystalline silicate/amorphous silicate ratio | analytical TEM, µXANES | keystone, u/m sections | common |
| | Isotopic anomalies, e.g., $\delta^{50}$Ti vs $\delta^{54}$Cr, $\delta^{50}$Ti vs $\delta^{46}$Ti, $\delta^{62}$Ni vs $\delta^{54}$Cr in coarse components | RIMS | potted butts | rare |
| | High-precision O-isotope composition of large silicates | SIMS | potted butts | common |
| | Evolution of carbonaceous material | STXM | u/m sections | rare |

**Summary of objectives**

- **Extend high precision O-isotope measurements to micrometer-sized samples**
- **Improve statistics of O isotope measurements of Wild 2 CAIs and chondrule fragments, to compare with corresponding meteoritic measurements**
- **Compare bulk O-isotopic composition of Wild 2 to planetary values**
- **Determine O-isotopic composition of Wild 2 fines**
- **Determine abundance of $^{16}$O-poor materials like cosmic symplectite in Wild 2**
- **Determine or constrain oxygen fugacity and formation temperature of Wild 2 rocks**
- **Compare minor element concentrations in olivines from Wild 2 with olivines from CP-IDPs and chondrite matrix**
- **Determine accurately crystalline silicate fraction and variability within Wild 2 samples**
- **Determine mean and distribution of low-Ca pyroxene to olivine ratio**
- **Search for isotopic anomalies, e.g., $\delta^{50}$Ti vs $\delta^{54}$Cr, $\delta^{50}$Ti vs $\delta^{46}$Ti, $\delta^{62}$Ni vs $\delta^{54}$Cr in coarse components**
- **Compare the bulk composition of fines and coarse-grained component to CI with sufficient accuracy to test hypothesis of complementarity**

**Technology advances**

- **Improve sample preparation for high-precision O isotopic composition measurements**
- **Improve sample preparation for O isotopic composition measurement of fines**
- **Improve ion yield in ion mass spectrometry to allow high-precision O isotope measurements on ultramicromed slices**
- **Develop capability of synchrotron hard X-ray transition metal K-edge spectroscopy with <20 nm spatial resolution**
- **Develop capability of synchrotron soft X-ray transition metal L-edge spectroscopy with ~10 nm spatial resolution**
- **Develop capability of site location through TEM/ALCHEMI**
- **Improve precision of minor element isotopic composition (Ti, Cr, Ni) measurements**
- **Develop high count-rate X-ray detectors with excellent energy resolution (< 10 eV)**

In order to explain the presence of refractory and chondrule melt fragments in the Wild 2 samples models of radial drift in the disk plane or above it in disk winds have proved important. In a model for the solar nebula simulating evolution in surface density of grains due to diffusion, gas drag, and viscous flows, Ciesla (2007) concluded that outward transport of high-temperature materials naturally occurs around the midplane of turbulent protoplanetary disks. In this way material such as micro-chondrules, CAIs or their fragments could potentially be shifted from the inner Solar System to the comet forming environment 10-30 AU. In support of this type of model, the coagulation timescale to grow from µm-sized grains to cm-sized pebbles is expected to be short ($<10^3$ a) compared to the radial mixing timescale ($>10^4$ a, Lorek et al 2016).

In contrast to this, Hughes and Armitage (2010) produced 1-D disk models to investigate radial drift of particles in the protoplanetary disk, and found no model that would sustain outward transport for large particles (>mm), but smaller particles could still be transported. Recent observations (Wozniakiewicz et al., 2013 ApJ) of the size distribution of components of CP-IDPs imply that size-sorting by grain size, composition or density may have occurred during transport. However, it was found that even the smaller particles (<20 µm) resided in the outer Solar System for only a short period of time ($<10^6$ years). Thus, if the material that formed Kuiper Belt planetesimals is made of the outwardly transported fine grained chondritic material (~20 µm), then the Kuiper Belt planetesimals would have had to have rapidly accreted this material, before it returned to the inner Solar System.. Thus the mechanism of radial drift seems to remain uncertain, leaving open the question of whether the Wild 2 components formed in the inner or outer Solar System. This is likely to be a topic for much further study and modeling.

Oxygen isotopic composition of Wild 2 particles

The isotopic composition of oxygen varies throughout the Solar System with bulk planets and asteroids being depleted by about 6% in the dominant isotope $^{16}$O relative to the rare isotopes $^{17}$O and $^{18}$O, compared to the Sun, inferred from measurements of solar wind returned by the Genesis mission (McKeegan et al. 2011). This "planetary" isotope signature is thought to represent mixing between the $^{16}$O-rich bulk solar composition and a $^{16}$O-depleted reservoir, whose nature and origin are a matter of substantial debate.  O isotopic variations of different magnitudes are observed at various spatial scales in extraterrestrial materials and provide a variety of information about the origin and early evolution of the Solar System.  At the smallest scales, O-rich presolar grains that condensed around stars can be identified by their highly anomalous isotopic compositions that can differ from Solar System values by orders of magnitude (Zinner 2014). Percent-level variations are seen at the scale of individual components of meteorites and IDPs, reflecting both their origins and fractionation occurring during parent-body processing. For example, CAIs are typically close to the $^{16}$O-rich solar composition, while chondrules, IDPs, and chondrite matrix tend to be closer to the more $^{16}$O-poor planetary signature, and some rare materials (e.g., so-called "cosmic symplectite") have been identified in chondrites and IDPs  with much more extreme (10-20%) $^{16}$O depletions, perhaps reflecting the primordial $^{16}$O-poor nebular reservoir. At the scale of bulk asteroids and planets, permil differences in O-isotopic compositions are a highly diagnostic classification tool. Therefore, different questions may be asked based on the level of measurement precision.

Some key questions regarding O isotopes in Wild 2 samples include:
- What is the abundance of isotopically anomalous presolar grains in comets?
- Do Wild 2 CAIs and chondrule fragments have similar O isotopes to those in meteorites indicating a genetic relationship?
- Is the bulk O-isotopic composition of Wild 2 solids planetary-like?
- Can the Wild 2 fines' O-isotopic compositions constrain their origin? (discussed in Objective 7)
- What is the abundance of $^{16}$O-poor materials like cosmic symplectite in Wild 2?
- Are grains found in the Stardust interstellar collector indeed interstellar in origin? This question is discussed in Objective 9.

As with other Stardust analyses, O-isotopic analyses pose significant technical challenges, including the difficulty of extracting and mounting small samples for high-precision analysis and need to distinguish samples or regions of samples that were altered by impact from those that were not.  Nonetheless, a number of O-isotopic studies on Wild 2 samples have been carried out thus far with increasing sophistication in sample preparation and analytical methodology, all are based on secondary ion mass spectrometry (SIMS). NanoSIMS measurements on impact crater residues on Stardust Al foils have been used to identify a few O-rich presolar grains and constrain their abundance (McKeegan et al. 2006, Stadermann et al., 2008) though additional searches are desirable. Relatively low-precision (> 10 per mil)

SIMS measurements revealed the first Stardust CAI (Track 25 Inti) to be $^{16}$O-rich, similar to meteoritic CAIs (McKeegan et al. 2006). Higher-precision analyses of silicate grains larger than a few microns extracted from the aerogel collector, including many chondrule-like fragments, have found planetary-like signatures similar to those seen in coarse grained chondritic silicates (Nakashima *et al.* 2012, Ogliore *et al.* 2015). But numerous $^{16}$O-rich grains have also been seen including LIME olivines and even a relict forsteritic olivine in a chondrule fragment (*Nakamura* et al. 2008)  O isotopes in fines have been addressed in two ways. Ogliore et al (2015) pressed a compressed piece of aerogel containing sub-micron particles into In and used SIMS imaging methods to determine the isotopic composition of individual embedded grains. This study found a wider range of O isotopes than has been observed in other fine-grained planetary materials including chondrite matrices and IDPs (excluding presolar grains). Snead et al (2015) measured impact residues on Stardust Al foils to avoid the problem of aerogel contamination and found planetary-like isotope signatures for most, but possible evidence for a cosmic symplectite-like contribution in one crater.

Additional high-precision O-isotope analyses on Stardust samples are crucial for better addressing the questions outlined above. Most analyses have been made on "potted butts" left over from microtoming particles. However, current sample preparation methodologies (e.g., Westphal et al 2011, Ogliore et al 2015) are for the most part difficult, time-consuming and often require specialized equipment and training, so simpler methods are highly desirable.

**A particularly important goal is to extend the high precision O-isotope measurements to micrometer-sized samples for the simple reason that the Wild 2 collection contains only a limited number of larger (terminal grains) but a nearly unlimited number of smaller fragments that are found in some tracks by the thousands**. Measurements of ultramicrotomed sections, of order 100 nm thick, could also be useful since these are relatively plentiful, but the amount of material available is insufficient to make high-precision (< 2 per mil at 2σ) oxygen isotopic measurements without improvements in ion yields.  Finally, continued comparison of the aerogel and Al foil samples will also be necessary for extracting the maximal scientific impact from the collection.

Oxidation state as a probe of mixing and transport

Urey and Craig (1950) were the first to recognize through systematic analysis that the oxidation state of Fe varies between meteorite groups.   Oxygen fugacity may increase with heliocentric distance of formation, although the relationship is probably not monotonic or straightforward.   Nevertheless, measurement of the oxidation state of Fe and other transition metals provides constraints on the formation environment of the components of Wild 2, and any affinities with meteorite groups.

The oxidation state of Fe in Wild 2 is distinct from that of any recognized meteorite family, and is metal-rich as compared with CP-IDPs (Westphal *et al.* 2009) although the comet also contains a major abundance of Fe-rich olivines which must have formed under quite oxidizing conditions.  The difference with CP-IDPs, which appears not to be related to selection biases, capture effects in aerogel (for Wild 2 samples) or high-speed atmospheric entry (for CP-IDPs), might be understood if comets are stratigraphically heterogeneous: Stardust sampled only the near-surface of Wild 2, probably within a thermal skin-depth, while CP-IDPs may sample the entire volumes of disrupted Jupiter-family comets (Ogliore et al. 2012b, Nesvorny et al. 2010). This hypothesis may be consistent with the recent observations by the Rosetta mission that provide evidence for periodic, stratified accretion of comet 67P/Churyumov-Gerasimenko, a Jupiter-family comet like Wild 2 (Massironi et al. 2015). The hypothesis is also complicated by mass loss regions on Wild 2 that are well over 100 m deep and the possibility that the entire comet may have lost hundreds of meters of its surface in the solar-driven sculpting of its highly complex pitted surface (Brownlee *et al.* 2004).

In large Wild 2 particles, the distribution of iron content in olivine and the correlations of minor elements (MnO, CaO, $Cr_2O_3$) with olivine iron content seen in Stardust samples are not observed in any individual meteorite class. The distributions can be reproduced by a mixture of contributions from known meteorite families (Frank *et al.* 2014; Brownlee and Joswiak, 2015). However, the apparent overabundance of Type II over Type I micro-chondrules and chondrule fragments in the Stardust collection suggests that the distribution of Fe-bearing silicate minerals from the inner Solar System to the outer Solar System may not have been homogeneous. Assuming that these objects derive from the inner Solar System, these observations may be used to constrain the timing and selectivity of large-scale transport in the early Solar System. Studies to date suggest that the compositional dispersions in Wild 2 olivine are similar to olivines in some very primitive CP-IDPs.

A study of Ti valence states using electron energy loss spectroscopy with a transmission electron microscope showed that variable Ti oxidation states exist in low-Fe augite in nodules within and between single fine-grained Wild 2 CAIs (Joswiak *et al.* 2015). This suggests that the fine-grained cometary CAIs (from Wild 2 and a giant cluster particle) may have moved between local environments with differing $f_{O2}$ perhaps due to variable dust/gas ratios in a hot inner Solar System environment prior to transport to the comet-forming region. Similar variability in Ti oxidation states in low-Fe augite in type A CAIs in chondrites were reported in CAI interiors and rims (J. Simon *et al.* 2005; S. Simon *et al.* 2007). Earlier work on the mineral osbornite in the Wild 2 CAI Inti showed variable V content within single submicrometer-sized grains also suggestive of either dynamic nebular conditions or of movement between different spatial locations during formation (Chi et al. 2009). Interestingly, $f_{O2}$ values calculated in the same study from Ti valence states in Inti fassaitic pyroxene, which formed at lower temperatures and C/O ratios than osbornite, are consistent with formation in conditions like those for Solar System CAIs.

Ongoing advances in synchrotron X-ray spectroscopy promise to enable better determination of oxygen fugacity. High spatial resolution hard X-ray microprobes, such as the new hard X-ray beamline at the National Synchrotron Light Source II, will allow of K-edge spectroscopy of the transition metals Ti through Fe, allowing for determination of oxygen fugacity over the entire range of solar-system values. The development of L-edge spectroscopic techniques for transition metals, currently demonstrated with ~10 nm spatial resolution and the capability of analyzing ultramicrotomed sections, will require a combination of extensive analyses of standards combined with *ab initio* calculations. Site occupancy measurements using ALCHEMI (Atom Location by CHanneling Enhanced MIcroanalysis) in Transmission Electron Microscopy may further constrain oxygen fugacity and formation temperature (Gainsforth et al. 2014).

Crystallinity

While crystalline silicates have been observed in circumstellar environments (Molster and Kemper 2005), it is known that interstellar silicates are almost entirely amorphous (Kemper et al. (2004, 2005)), so the presence of crystalline silicates in comets implies that a substantial fraction of interstellar material was subject to high-temperature processes prior to incorporation into comets. An alternative explanation is that crystalline silicates in comets and also seen in circumstellar disks forms by low temperature annealing of amorphous materials. This is clearly not the origin of many Wild 2 crystalline materials, particularly those related to CAIs and chondrules but it is possible that materials with such origin do exist in the comet but are hard to positively identify. To the extent that high temperature processes may have been restricted to the inner Solar System, the fraction of crystalline silicates (Westphal *et al.* 2009, Stodolna *et al.* 2012a) and the abundance of semi-volatile elements, after correction for aerogel capture effects, may be sensitive quantitative indicators of large-scale nebular mixing, and may provide constraints that can be compared with models (e.g., Ciesla 2007). **Further studies are needed to improve over the current measurements of the crystalline silicate fraction, and to determine its variability among Wild 2 samples.**

Low-Ca pyroxene/olivine ratio

The ratio of low-Ca pyroxene to olivine varies among meteorite groups (Dobrica *et al.* 2009), from ~0.2 for CM meteorites to ~1 for CR meteorites. This number is only approximately known to be ~1 in Wild 2, but **a more accurate measurement of the mean and distribution of this ratio will help to constrain the mix of materials from which coarse particles in Wild 2 derive.**

Inherited nucleosynthetic anomalies

Different meteorite groups show subtle, systematic isotopic anomalies. For example, anomalies in $^{46}$Ti, $^{50}$Ti, $^{54}$Cr, and $^{62}$Ni have been interpreted as nucleosynthetic anomalies inherited from distinct Solar System reservoirs (Trinquier *et al.* 2009). $^{54}$Cr is also correlated with Cu/Yb and Ga/Yb ratios, and varies systematically among carbonaceous chondrites, with low $^{54}$Cr and low Cu/Yb and Ga/Yb for CO meteorites and high values for CI chondrites. The isotopic anomalies are extremely small, less than 1 part in $10^5$, but a next-generation, high-yield instrument (e.g., RIMS, Stephan et al. 2016) may be able to approach this precision on ≥20μm particles. The elemental ratios might be measurable with high energy resolution X-ray detectors such as micro-calorimeters (Silver *et al.* 2014): the challenge will not be in measuring the abundances in the numerators (Cu and Ga) but in the denominator (Yb).  Current micro-calorimeters are limited by low maximum count rates.

Compositional complementarity

If coarse-grained materials in Wild 2 (micro-CAIs, micro-chondrules, etc.) are indeed xenoliths derived from the inner Solar System, then their composition may not be complementary to that of the fine-grained material, which, if it is very primitive, may have an elemental composition similar to that of the Sun.   Coarse-grained particles systematically have larger ranges in the aerogel than fine-grained materials, which tend to stop quickly and so are found in the bulbous regions of tracks.  The separation is not perfect –  some fine-grained material that is mechanically attached to large particles can penetrate to large depths in the aerogel – but there is sufficient separation that compositional differences can be explored in principle.  **The limited synchrotron X-ray analyses (Flynn *et al.* 2006, Ishii *et al.* 2008b, Lanzirotti *et al.* 2008, Schmitz *et al.* 2009, Tsuchiyama *et al.* 2009) are not yet sufficient to test this hypothesis, but future efforts with improved statistics in a coordinated effort may be able to do so**.  Mobilization of volatile components in aerogel add to the challenge of this measurement.  On the other hand, if the coarse-grained materials are complementary in elemental composition to the fines, so that the total composition of Wild 2 is near solar, then this may be an indication that the micro-chondrules and micro-CAIs formed *in situ* in the outer nebula, as suggested by Bridges et al. (2012), or that the fines were transported from the inner solar system also.

**Objective 3: Determine age of cometary material relative to CAIs**

| Question and Objective | measurement | instrument | sample requirements | sample frequency |
|---|---|---|---|---|
| 3 When do comets form in the construction sequence of the Solar System? *Determine relative and absolute age of cometary materials relative to CAIs, and timing of large-scale transport and mixing* | $^{26}$Al age (Al-Mg systematics) | SIMS, RIMS | high Al/Mg plagioclase | rare |
| | $^{53}$Mn age (Mn-Cr systematics) | SIMS, RIMS | high Mn/Cr sphalerite | rare |
| | $^{87}$Rb age (Rb-Sr systematics) | RIMS | high Rb/Sr phases (sulfide?) | rare? |

**Summary of objectives**

- **Search for evidence of live $^{26}$Al in large numbers of well-suited Wild 2 samples**
- **Search for evidence of live $^{53}$Mn in large numbers of well-suited Wild 2 samples**
- **Measure age of Wild 2 samples with long-lived radioisotope systems (e.g., Rb-Sr)**

**Technology advances**

- **Develop improved synchrotron microprobe search capability for targets for Al-rich, Mg-poor phases**
- **Improve methods and standards for Al-Mg and Mn-Cr isotope measurements in Wild 2 samples**
- **Develop capability of using long-lived clocks to measure absolute age of Wild 2 materials**

Measurements of short-lived and long-lived radionuclides and their decay products in meteorites are very useful to determine the chronology of events in the Solar System. The radioactive parent of short-lived systems used in cosmochemistry, such as $^{26}$Al-$^{26}$Mg and $^{53}$Mn-$^{53}$Cr, was assumed to be present in the nascent solar nebula at some uniform abundance. Using a sample that has been suitably dated for both a long-lived and short-lived system, the short-lived system is "anchored" to the long-lived system and an absolute date can be inferred. The disadvantage of short-lived systems is that their chronological information is dependent both on the anchor and the assumption that the initial abundances were the same everywhere in the young Solar System. The advantage is that certain mineral phases can concentrate a radioactive parent element with respect to the daughter element. The initial amount of radioactive parent isotope in a sample (which has long since decayed) is measured by determining the proportionality constant between the radioactive parent element and the excess daughter isotope.

$^{26}$Al decays to $^{26}$Mg with a half-life of 730,000 years. Because of this short half-life, the presence of $^{26}$Al in early solar-system materials can only be inferred through a detection of an excess of its daughter, $^{26}$Mg. A search for excess $^{26}$Mg requires samples with very large Al/Mg ratios, and has been carried out on only four Wild 2 particles so far: two CAI fragments ("Inti", Ishii *et al.* 2010, "Coki", Matzel *et al.* 2010a), one chondrule fragment ("Iris", Ogliore *et al.* 2012b) and one ferromagnesian silicate ("Pixie", Nakashima *et al.* 2015). No evidence for live $^{26}$Al was found in any of these samples. Coki has mineralogy similar to a type C CAIs in meteorites, of which half are likely to have been remelted in the chondrule-forming region (MacPherson 2014). The upper bound of $1 \times 10^{-5}$ inferred by Matzel et al. 2010 for the initial $^{26}$Al/$^{27}$Al ratio in Coki is consistent with the low abundance of $^{26}$Al in type C CAIs (Krot *et al.* 2005). On the other hand, the mineralogy and $^{16}$O-rich composition of Inti indicate that this CAI was not melted during chondrule formation.

A challenge with the Al-Mg dating system is the lack of Al-rich, low-Mg phases, such as anorthite and hibonite, in Wild 2 samples. Even after years of observation over dozens of tracks and hundreds of fragments by numerous labs around the world, we still have not found an ideal fragment for a best-case Al-Mg study. **An important future contribution will be to develop the capability of efficiently searching for Al-rich, Mg-poor phases, probably using advanced synchrotron X-ray microprobes.**

In one model of the early Solar System, the distribution of $^{26}$Al was heterogeneous but was later homogenized in the protoplanetary disk by radial mixing (Krot *et al.* 2012). Chondrules would have formed after this homogenization, so dating their formation using the Al-Mg system is less problematic than dating refractory inclusions like Inti and Coki. The chondrule fragment Iris (Ogliore *et al.* 2012b) and the ferromagnesian silicate Pyxie (Nakashima *et al.* 2015) did not show evidence for the former presence of $^{26}$Al (inferred $^{26}$Al/$^{27}$Al< $3\text{-}4 \times 10^{-6}$). Assuming $^{26}$Al was homogenized at the canonical CAI value of $^{26}$Al/$^{27}$Al $= 5 \times 10^{-5}$ in the protoplanetary disk soon after the formation of CAIs (MacPherson 2005), these grains must have formed more than 3 Ma after CAIs. If Pyxie and Iris formed in the inner Solar System, this implies that large-scale radial transport had to occur very late in the young Solar System in order to move these particles out to the trans-Neptunian region for incorporation into comet Wild 2. The formation of Jupiter effectively cuts off in-disk transport of material, so the presence of these late-forming inner Solar System objects in comet Wild 2 has far-reaching implications for planet formation (Ogliore *et al.* 2012b). An additional complication is the possibility of delayed injection of $^{26}$Al in the Solar System. Many Al-rich refractory inclusions in meteorites show no evidence of radiogenic $^{26}$Mg, which has been suggested by some as indicating not a late formation but rather formation before $^{26}$Al was introduced into the inner nebula (MacPherson 2005, Liu et al 2009; Koop et al 2016). This might happen if $^{26}$Al was injected (e.g., by a supernova) into the collapsing cloud or an already-formed disk where some $^{26}$Al-free refractory objects had already formed.

The $^{53}$Mn-$^{53}$Cr system has been used to measure formation ages of minerals formed by aqueous alteration on meteorite parent bodies. Secondary minerals like carbonates and fayalites tend to concentrate Mn and exclude Cr, making them ideal candidates for Mn-Cr dating assuming that they formed during the lifetime (half-life=3.7 Ma) of $^{53}$Mn. One Wild 2 particle, a fayalite-silica intergrowth named Ada, has been measured for the Mn-Cr system by two different labs (Ogliore *et al.* 2014, Matzel *et al.* 2014), but the measured Mn/Cr ratios were too low to obtain meaningful constraints on the initial $^{53}$Mn abundance. However, similar objects in ordinary chondrites are interpreted to be formed by aqueous alteration. Thus, if Ada (or similar grains in the Wild 2 collection) formed by a similar process, **more robust Mn-Cr measurements could provide useful constraints on the timing of its formation relative to alteration timescales in asteroids**, if $^{53}$Mn in the early Solar System was homogeneous.

Progress in assessing the Al-Mg and Mn-Cr systematics in Wild 2 samples will also benefit from recent technical advances. For example, a new high-spatial resolution O- ion source for the NanoSIMS ion microprobe (Matzel et al. 2014) allows measurement of smaller samples with less contribution from surrounding materials than previously possible. Precise interpretation of SIMS data for the Mn-Cr system – for meteorite samples as well as for Wild 2 dust – has also been hampered by the difficulty of preparing accurate standards for determining Mn/Cr elemental ratios. There has been recent progress on this front (e.g. Jilly et al. 2014), but additional development of improved standards is highly desirable.

Aside from pre-solar grains, which are known to predate the Solar System, the oldest Solar System materials that have been reliably dated are the CAIs. It is assumed that cometary solids formed no earlier than this, but actual evidence of this is lacking. Long-lived systems, such as Pb-Pb dating, are powerful as they provide an absolute age since closure of the radioisotope system (e.g., Connelly *et al.* 2012). However, the relatively low cosmic abundances of Pb, U, and other elements in long-lived systems means samples larger than ~20 mg are required. Samples from comet Wild 2 are too small to be dated using long-lived systems with current techniques (thermal ionization mass spectrometry and high-resolution inductively coupled plasma source mass spectrometry). **A major goal of current and future analytical developments is to achieve sufficient precision that long-lived systems in Wild 2 samples can be made accessible.** The newest generation RIMS (resonance ionization mass spectrometry) instrument, CHILI (Chicago Instrument for Laser Ionization; Stephan et al., 2016), is designed to determine absolute ages in small grains with unprecedented precision. Rb-Sr dating should be possible with ≥20 μm particles.

**Objective 4: Determine character and origin of organic matter**

| Question and Objective | measurement | instrument | sample requirements | sample frequency |
|---|---|---|---|---|
| 4 How and where do cometary organics form? *Determine origin of organics* | H, C, N isotopes of organics | nanoSIMS | shielded material (possibly behind terminal particles), craters in foils | rare |
| | Composition and structure of organics | Raman, STXM, XANES, L$^2$MS, GCMS | shielded material (possibly behind terminal particles) | rare |

**Summary of objectives**

- **Determine affinities of Wild 2 organics with organics in chondrites and/or IDPs through analysis of more samples**
- **Reveal organic reaction networks that may have generated complex organic matter on the Wild 2 body**
- **Substantially increase statistics of identification and characterization of isotopically anomalous organic matter**

**Technology advances**

- **Development of new sample-preparation media that do not solubilize organics or penetrate particles/silica aerogel**
- **Development of *ultra*-L$^2$MS for analysis of organics with <5 μm spatial resolution**
- **Development of nanoFTIR for analysis of organics with <50 nm spatial resolution**
- **Utilization of direct-electron detectors, 40-60 keV electron sources, and aberration-corrected lenses for high resolution (S)TEM observations at low beam damage conditions.**

Although organic matter is famously known to be abundant in cometary gas and on cometary surfaces and in coma particles (Goesmann et al. 2015; Greenberg 1998; Lawler and Brownlee 1992), the collection and return of pristine cometary organics was only a secondary goal of the Stardust mission. Thus, the discovery of carbonaceous matter preserved alongside mineral grains in captured terminal particles was an unexpected bonus of the mission (Brownlee et al. 2006; Sandford et al. 2006). Cometary carbonaceous material could have been a major component of the Earth's original inventory of prebiotic carbon (Pasek and Lauretta 2008). Three general categories of carbonaceous matter were found in the Stardust aerogel and foils: (1) soluble compounds present in the coma that infiltrated the aerogel tiles, (2) soluble and labile organic matter present in coma aggregate grains, and (3) insoluble, macromolecular organic matter. Cometary soluble compounds impinging on the collector travelled through the aerogel, ultimately depositing on the Al foil surfaces lining each aerogel tile. After exposure to hot water extraction and acid vapor hydrolysis, the amino acid glycine was detected from the side foils, with a "heavy" C isotopic composition, indicating a cometary, rather than terrestrial, origin (Elsila et al. 2009). α- and β-alanine were also detected, but in concentrations too low for a robust measurement of their isotopic composition.

Less-soluble, labile organic matter has been observed as extensive organic halos surrounding some particle capture tracks (Bajt et al. 2009), high concentrations of organics within densified aerogel without a terminal grain (De Gregorio et al. 2011), and soluble organic components that have been extracted from fine-grained terminal particles after embedding in epoxy (Cody et al. 2008a), but no robust characterization of isolated soluble cometary material has been performed. Cometary insoluble organic matter (IOM), on the other hand, is rare but can be found both in fine-grained terminal particles, isolated carbonaceous grains embedded in the walls of bulbous Type B tracks, and as carbonaceous terminal grains. Coherent fragments of cometary IOM have been identified by TEM and/or STXM in thirteen Stardust particles. In nine of these cases, the IOM contains a consistent functional group chemistry, dominated by carboxyl and ketone bonding, with a variable amount of aromatic C=C bonding (e.g., De Gregorio et al. 2011). This functional group profile is similar to what is observed in IOM from primitive carbonaceous chondrites (Cody et al. 2008b) and IDPs (Keller et al. 2004), but without the variability in aromatic carbon abundance observed in the cometary samples. The lowest aromatic carbon abundance is reported from the terminal particle of Track 57 ("Febo"; Matrajt et al. 2008) and a bulb wall grain from Track 80 (De Gregorio et al. 2014), while the highest aromatic carbon abundance was found in carbonaceous nanoglobules (De Gregorio et al. 2010, 2011). Two of the carbonaceous grains are composed of poorly graphitized carbon (Matrajt et al. 2013b; De Gregorio et al. 2014), indicating that they were subject to a heating event, which is unexpected in a cold outer Solar System body. Graphitization could have occurred by flash heating of cometary organics during hypervelocity capture in aerogel, but it is also possible that the poorly graphitized carbon formed within the early solar nebula, prior to accretion onto Wild 2. The functional chemistry of the two remaining verified carbonaceous grains has not been characterized. Polycyclic Aromatic Hydrocarbons (PAHs) have also been observed in craters in the Stardust Al foils (Leitner et al. 2008)

Identification and characterization of cometary organic matter is complicated by intrinsic organic contaminants present in the silica aerogel as well as the potential for destruction or alteration from flash heating during particle capture (Fries et al. 2009). A significant research effort was undertaken to characterize possible aerogel contaminants by a variety of techniques (Sandford et al. 2010). All research programs that target cometary organics must now take into account these intrinsic contaminants in addition to those introduced by the terrestrial environment and sample preparation. In addition, the existence of soluble cometary organics in the Stardust collection requires that some kind of nanoscale imaging (often TEM) be included in order to distinguish organic-rich aerogel from coherent carbonaceous grains. An example of this can be seen by comparing the initial preliminary examination results on cometary organic matter (Sandford et al. 2006) to a later reanalysis of the same samples by TEM (De Gregorio et al. 2011). The eight carbonaceous grains described in the initial 2006 study showed a wide range of organic chemistry. However, the 2011 reanalysis showed that only one grain represented verifiable, authentic, unaltered, cometary organic matter (with two others not available for reanalysis). The apparent wide range of organic chemistry was reduced to a single functional group profile similar to that seen in chondritic IOM and IDPs (De Gregorio et al. 2011).

Meteoritic and IDP organics are commonly characterized by searching for isotopic anomalies in H, N, and, more rarely, C (typically, enrichments in D, $^{15}$N, and/or $^{12}$C). The origin of these isotopic signatures is hotly debated but they may indicate an origin in the Sun's parental molecular cloud or the outermost regions of the early solar nebula. In any case, the presence of such anomalies in Wild 2 grains is additional evidence of a cometary origin, similar to the methodological approach used to verify cometary glycine on Stardust Al foils (Elsila et al. 2009). Isotopic analysis of Wild 2 carbonaceous grains have revealed mostly terrestrial-like D/H ratios, though a few D-enrichments, with $\delta$D up to ~2,000 permil, have been observed (McKeegan 2006; Matrajt 2008). Nitrogen is mostly terrestrial-like as well, but $^{15}$N-rich carbonaceous grains have been identified both in isolated particles from along bulbous tracks (McKeegan 2006; De Gregorio 2010) and from shielded fine-grained material in terminal particles (Matrajt 2008), with $\delta^{15}$N up to ~1200 permil. The range of N isotopic ratios is slightly more limited than that seen in carbonaceous chondrites and IDPs, whereas the degree of D enrichment is much more modest. Some carbonaceous meteorite classes (particularly CM, CV, and CO chondrites) have similar ranges of $\delta^{15}$N and $\delta$D values (Alexander et al. 2007). These meteorites also tend to have fewer "hotspots" with extreme isotopic enrichments, which may be consistent with the current Wild 2 isotope data. However, these isotopic compositions may not be so surprising given the fact that so much of the mineral component in the Stardust collection has inner Solar System origin, and that conditions on the comet are not conducive to the *in situ* formation of IOM. **Transient aqueous fluids on Wild 2 (Miles and Faillace 2012) could generate the not insignificant amounts of soluble organics that have been observed so far, making the isolation of soluble organic matter from Stardust terminal particles and aerogel an essential goal for future work.**

A few studies have directly investigated the possible genetic relationship between the carbonaceous components of Wild 2 particles and those found in primitive carbonaceous chondrites and IDPs. Using STXM, Wirick et al. (2009) found that chondritic IOM contains generally more aromatic functionality than that of Wild 2 particles and IDPs. That IOM is altered by the extraction process is a complication (Flynn et al 2010). In addition, the Wild 2 carbonaceous matter showed a range of functional group diversity that mimicked that observed in the IDP samples, although, as discussed above, some of this diversity could be due to contaminants and interactions with silica aerogel during hypervelocity capture.

Comparisons between nanoscale morphology and textures of carbonaceous matter in Wild 2 particles and IDPs and found striking similarities (Matrajt et al. 2013a). Wild 2 carbonaceous matter typically shows a dense, coherent, "smooth" texture, but globular forms, speckled "dirty" textures, and lacy or vesicular "spongy" textures have also been observed (Matrajt et al. 2013a, 2013b), all of which are found in IDPs (Matrajt et al. 2012). In contrast, chondritic IOM tends to be dominated by a porous, mesoparticulate, "fluffy" texture, along with globular forms and larger "smooth" carbonaceous grains. While these studies suggest a direct link between IDPs and ejected comet dust, it would be improper to exclude a genetic link with chondritic IOM, keeping in mind that even the most unaltered, petrologic type 3.0 chondrites have still experienced some minor amount of parent body heating, pressure, and aqueous fluid flow, which can change and homogenize the functional group profile of accreted organic matter. In addition, carbonaceous matter in IDPs is mainly observed as organic coatings on mineral grains, and only a single example of an organic-coated grain has been reported from the Stardust collection (Matrajt et al. 2008). **The question of potential affinity of Wild 2 organic content with that found in chondrites or IDPs is certainly ambiguous at present, at least in terms of the organic content of Wild 2, and the most obvious remedy is to gather more observations of carbonaceous cometary samples.**

A corollary question, which is very much unaddressed, is that if Wild 2 carbonaceous matter is genetically related to organic matter that is prevalent in IDPs or carbonaceous chondrites, why do we not find more of it in the Stardust collection? Three factors may be responsible for the apparent lack of carbonaceous samples. First, carbonaceous matter may simply not be abundant in the coma dust that was exposed to the Stardust collectors. Polarimetric observations indicate that the polarization induced by coma dust in the circumnucleus halo region of Wild 2 is inconsistent with the high refractive indices of solid carbonaceous particles, and therefore, the abundance of carbonaceous matter in the coma region, from which the Stardust spacecraft collected material, is naturally low (Zubko et al. 2012). Secondly, flash heating and explosive volatilization associated with hypervelocity capture into silica aerogel may have caused a significant sampling bias against carbonaceous matter. Finally there may be a sample selection bias due to traditional sample preparation techniques that are not appropriate for analysis of organic samples. All of the carbonaceous Wild 2 samples that have been verified as cometary (as opposed to contaminants or organic-bearing silica aerogel) have been prepared by alternative methods--embedding in molten S (e.g., Cody et al. 2008a), which can be removed by gentle heating, or acrylic, which can be dissolved with chloroform vapor (e.g., Matrajt et al. 2008). Similarly, placing prepared samples on a C-free substrate for analysis will greatly increase the ease of discovering cometary carbonaceous matter. However, it should be noted that for a large survey of 16 Stardust tracks prepared by the acrylic method, carbonaceous cometary material was only found in five of them.

*Technological Advances Related to Organic Analysis*

It is not surprising that much of the research on organic matter in Wild 2 has utilized TEM and STXM techniques, as these two instruments gather information on organic bonding with the necessary nanoscale resolution to separate authentic cometary material from silica aerogel or terrestrial contaminants. However, several technological developments within the last ten years may prove extremely useful for future organic analysis of Stardust samples. New instrumentation for existing TEM capabilities include direct electron detectors, the availability of aberration-correcting lens systems, and low-energy 40-60 keV electron sources, which provide better spatial and spectral resolution while significantly lowering the required electron dose delivered to the sample. These advancements, then, allow for better analyses with a marked decrease in beam damage in sensitive organic samples.

A more complete picture of organic matter in Wild 2 can be gained by using additional spectroscopic techniques, such as laser microprobe mass spectrometry and FTIR, which could provide complementary information to TEM and STXM analysis. Unfortunately, the spatial resolution of these alternative techniques is often too poor to be robustly applied to carbonaceous matter in Stardust samples. However, a unique two-step, laser desorption laser ionization mass spectrometer ($L^2MS$) has been in continual development since the Stardust samples were returned (Clemett et al. 2010). The *ultra*-$L^2MS$ can probe both aliphatic and aromatic bonding with a 10 μm beam spot, and recent improvements to the instrument have increased its sensitivity by a hundred-fold over previously-published work. **Novel, NASA-funded instrumentation, such as *ultra*-$L^2MS$,** may be promising for revealing the functional chemistry of soluble cometary organics that have penetrated into the silica aerogel during capture (Clemett et al. 2014).

The high lateral resolution and chemical specificity provided by NanoFTIR is an emerging technique that may have applicability to understanding the origin and nature of cometary organics and the role that aqueous alteration may have played on the cometary parent body. NASA funding of instrument development has allowed for the recent demonstration that atomic force microscope (AFM) tip-enhanced spectroscopy can be used to construct infrared spectral maps of returned samples with <50 nm lateral resolutions. Recently, NanoFTIR mapping has been used to construct infrared spectral maps of the Wild 2 grain Iris (Dominguez *et al.* 2014) and Caligula (Gainsforth *et al.* 2013b). This work, so far, has shown that the technique is capable of identifying silicate minerals and mapping out mineralogical heterogeneity down to 50 nm, and recent work on Murchison indicates that the resolving power of NanoFTIR could be applied to identification and mapping of important prebiotic organics (Dominguez *et al.* 2015).

Commercial NanoFTIR instruments are now starting to become available, and a NanoFTIR instrument has recently been installed at the Advanced Light Source synchrotron. Looking towards the future, there are several challenges and opportunities for the application of AFM-tip enhanced spectroscopy of planetary materials. First, the technique relies on the availability of IR laser illumination sources, which are expensive and currently limited in wavelength coverage. However, the availability of IR lasers (e.g., Quantum Cascade Lasers) is continuously expanding and it seems likely that sources will soon emerge with wavelengths that allow for the in-situ imaging of additional important functional groups including -OH and $H_2O$. In addition, coherent THz illumination sources now exist that should provide sensitivity to astrophysically significant carbon and hydrogen bearing compounds such as polycyclic aromatic hydrocarbons (PAHs). Efforts to apply sub-micron scale IR and THz spectral mapping should facilitate further progress in this area and help answer some of the unresolved questions highlighted in this document.

**Objective 5: Search for signs of exposure to liquid water**

| Question and Objective | measurement | instrument | sample requirements | sample frequency |
|---|---|---|---|---|
| **5 Are products of aqueous alteration products present in comets?** *Establish if hydration can occur on cometary parent bodies and if liquid water is present.* | Concentration of aqueous alteration products (phyllosilicates, carbonates, cubanite, pentlandite, magnetite) | analytical TEM, µXRF, NanoFTIR, micro-Raman, XRD, XANES | shielded material (possibly behind terminal particles), craters in foils | rare |

**Summary of objectives**

- **Determine whether carbonates formed by aqueous processes can be distinguished from carbonates formed by nebular processes**
- **Determine the distribution of concentration of carbonate contamination in aerogel, and in aerogel tiles near Wild 2 tracks with reported carbonates**
- **Determine whether sulfides formed by aqueous processes can be distinguished from sulfides formed by nebular or other processes**
- **Determine whether magnetite formed by aqueous processes can be distinguished from magnetite formed by nebular or other processes**

**Technology advances**

- **Development of FTIR and nanoFTIR for analysis of hydrated phases**
- **Development of ability to measure H concentration using highly monochromated TEM/EELS**

Phyllosilicates

Classic aqueous alteration products such as phyllosilicates should exist in Wild 2 if alteration occurred inside the parent body with any substantial frequency. Although they have not been found, it is possible that they might exist or have existed in the bulbous regions of tracks or in crater residues. Even if they all melted during capture, their compositions could have left compositional clues to their former presence (Wozniakiewicz et al. 2015), although the presence of other minerals in the impacting particles could complicate the search for such signatures.. Smectite, serpentine, and tochilinite in carbonaceous chondrites and hydrated IDPs have distinctive elemental abundances that differ from chondritic values. These, along with other possible clues such as vesicles, could provide evidence for the presence of fine hydrated matter that melted during capture. Further, large grains (more than ~2 um), if present, should have had residual TEM-visible structures remaining as shown by Noguchi et al. (2007) in laboratory-based hypervelocity capture experiments and subsequent TEM, synchrotron XRD, and FE-SEM analyses. This kind of search could reasonably be done in sections from bulbous tracks. Other clues may be provided by searching for organic molecules associated with aqueous alteration, perhaps within the largest terminal particles that are less susceptible to capture alteration effects.  **In principle, NanoFTIR (discussed in more detail under Objective 5) or similar tip-enhanced IR imaging methods may be used to map surface bound water on samples and to identify products of aqueous alteration.**   (Dominguez *et al.* 2014; Dominguez *et al.* 2015). **Ishii et al. (2016) have recently demonstrated the capability of detecting H by TEM/EELS, which holds the promise of measuring this major element at high spatial resolution (several nm) for the first time in Wild 2 samples.**

It is worth also stressing that in all impact experiments using aerogel, we see a wide range of heating behavior.   If phyllosilicates had been a significant phase we would have seen some tell tale morphologies of heated phyllosilicates, as we see in agglutinates in C chondrites (Zolensky et al., 2015) and thermally metamorphosed C chondrites (Tonui et al., 2014).

Carbonates

Carbonates are common products of aqueous alteration, so those found in extraterrestrial samples are commonly taken as a signature of the presence of liquid water.  Burchell et al. (2006) showed that carbonates can survive capture into aerogel.  While no large carbonate grains have yet been identified in returned Wild 2 samples, numerous submicron carbonate grains, frequently intermixed with other Wild 2 mineral grains, have been found (Flynn *et al.* 2009), including Mg-Fe-carbonates associated with amorphous silica and iron sulfides (Mikouchi *et al.* 2007).  Nano-scale carbonates also found assoiuated with glass and D-rich IOM in highly primitive GS-IDPs (Busemann et al 2009).   A concern until recently has been that these carbonates may be a contaminant from the aerogel, but a recent search for carbonates in track-free Stardust flight aerogel by George Flynn and colleagues detected no carbonate (G. Flynn, private communication). This showed that the aerogel capture medium is low in carbonates, to the extent that they are homogeneously distributed, so these submicron carbonates are likely to be indigenous to Wild 2 (Flynn, private communication).  .  Further, while Ca carbonate could plausibly be a manufacture contaminant in aerogel, Mg carbonates are unlikely (Mikouchi et al. 2007).  Because carbonates are readily destroyed during aerogel capture, the implication is that Wild 2 may contain abundant submicrometer-sized carbonates.  Carbonates can also in principle be formed without the presence of liquid water, in gas-phase reactions in the nebula (Toppani *et al.* 2005), so the presence of carbonates is not an unambiguous signature of aqueous alteration.  Whether or not carbonates formed by the two processes can be distinguished by more detailed analyses is not yet known.  Further experimental petrology experiments, to attempt synthesis of carbonates under a suite of distinct conditions, could elucidate diagnostic criteria for origin locations and processes.

Sulfides

Sulfides in Wild 2 are predominantly pyrrhotite $Fe_{(1-x)}S$ with near troilite FeS composition, but the existence of a number of unusual sulfides implies diverse origins.  Based on a synchrotron survey of 194 grains in 11 Stardust tracks, the S/Fe ratio is ~0.3 and most of the S is bound into pyrrhotite/troilite (Westphal *et al.*, 2009).  Some pyrrhotites are terminal particles several micrometers in diameter while others are located within assemblages with igneous textures (Joswiak *et al.* 2012, Gainsforth *et al.* 2013b).  In addition, pyrrhotite can occur in association with pentlandite $(FeNi)_9S_8$, sphalerite ZnS, and kamacite FeNi alloy in igneous textures, though they likely did not form near equilibrium (Joswiak *et al.* 2012).  Other nano-sulfides have euhedral morphologies and compositions that suggest they formed in a nebular environment (Stodolna *et al.* 2014, Gainsforth *et al.* 2014).  The presence of cubanite $CuFe_2S_3$ has been interpreted as evidence for aqueous processing (Berger *et al.* 2011), although it is known that cubanite can be formed in non-aqueous environments,   The presence of pentlandite, a common component of hydrated IDPs, is consistent with aqueous alteration, but recently it has been demonstrated that pentlandite can form by other mechanisms (Schrader *et al.* 2015). Even if cubanite and pentlandite indicate aqueous activity on Wild 2, Wild-2 sulfides plot within the Fe-Ni-S ternary plot as two modes: either pyrrhotite/troilite, or pentlandite, with few if any compositions between (Zolensky *et al.* 2006).  This limits the extent of aqueous processing since by comparison to hydrous IDPs hydrous processes produce a wide range of pentlandite compositions.   It is clear that sulfides trace a very diverse range of formation and processing environments and perhaps provide some of the most direct evidence for aqueous alteration within the Stardust sample suite, while at the same time, limiting the scope of parent body alteration.

Magnetite

The iron oxide magnetite has been identified with synchrotron X-ray diffraction and X-ray Absorption Near-Edge Spectroscopy (Fe K XANES) in Stardust samples (Changela et al. 2012, Bridges et al. 2015).  One terminal particle from track 183 was found to consist mainly of Cr-rich magnetite nanoparticles (De Gregorio et al. *submitted*). These detections of magnetite are significant because in carbonaceous chondrites Ti-free magnetite is a typical product of hydrous alteration of ferromagnesian silicates like olivine and pyroxene.  The origin of magnetite in the matrix of carbonaceous chondrites through reaction of ferromagnesian silicates with parent body water was established by Kerridge et al. (1979) in a study of CI chondrites and provides a clue for ongoing research that materials in Wild 2 did experience significant aqueous alteration.  On the other hand, high-temperature heating of chondritic metal, perhaps during nebular shocks or chondrule-forming events, can also produce magnetite and chromite with trace metal abundances proportional to that of the original metal (Lauretta and Schmidt 2009), and it is possible that magnetite could even form as a primary nebular condensate within specific P-T-$f_{O2}$ conditions. More detailed observation of Wild 2 magnetite is clearly necessary in order to reliably assess their origin.

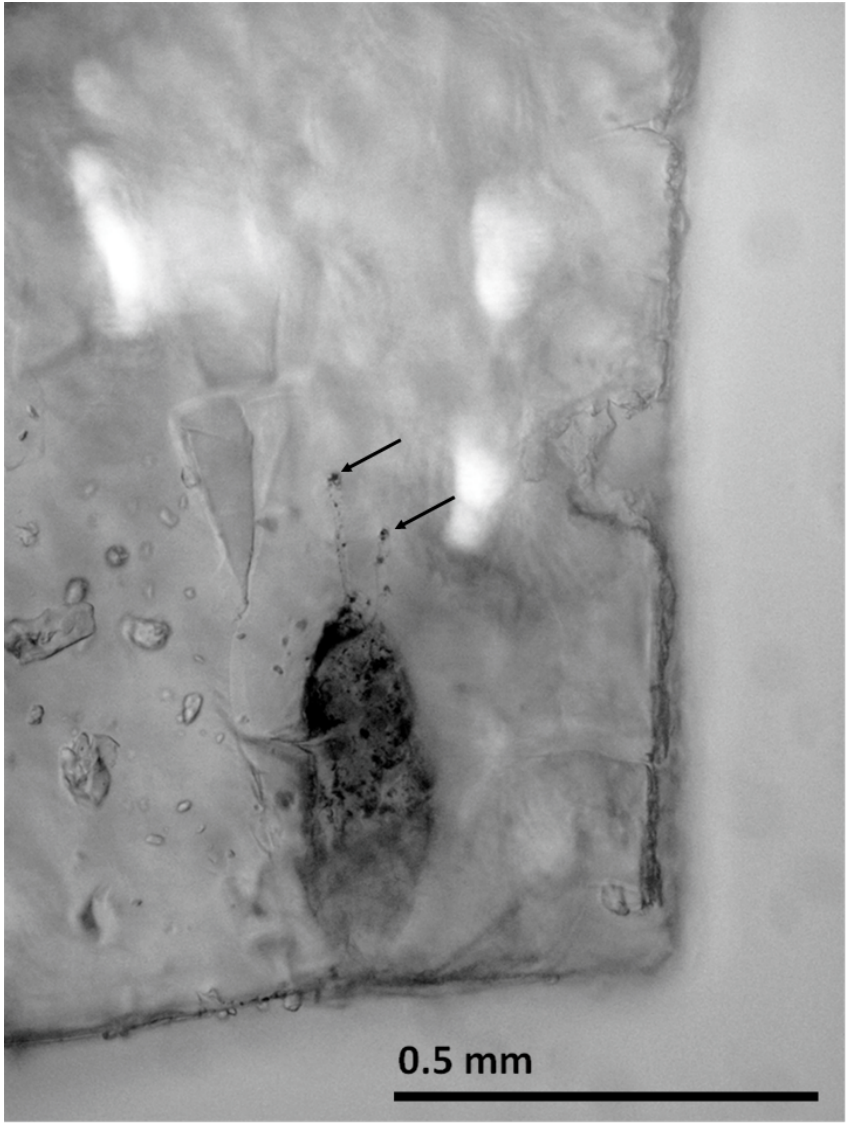

Fig. 6. Two magnetite terminal particles in Stardust track 187, identified by XRD at the Diamond synchrotron (Bridges et al. 2015).

**Objective 6: Determine formation conditions of micro-chondrule/chondrules, micro-CAIs, and metals**

| Question and Objective | measurement | instrument | sample requirements | sample frequency |
|---|---|---|---|---|
| **6 How and where do cometary rocks form?** *Determine micro-chondrule and micro-CAI formation conditions* | Element partition in microchondrules, and micro-CAI volatile element concentration | SEM, analytical TEM | microchondrules | common |
| | Ni et al. trace siderophile concentrations in metal grains | siderophile-free sample preparation, LA-ICP-MS | large metal particles (e.g., Simeio) | rare |

**Summary of objectives**

- **Substantially improve statistics of detailed elemental, mineralogical, and isotopic analyses of microchondrules/chondrules**
- **Substantially improve statistics of detailed elemental, mineralogical, and trace-element analyses of microCAIs**
- **Substantially improve statistics of detailed elemental, mineralogical, and trace-element analyses of metals**

**Technology advances**

- **Development of W-, Pt-, Au-, Ga-free FIB techniques for sample preparation of metal particles**
- **Develop efficient search techniques for micro-CAIs using synchrotron X-ray microprobes**

Micro-chondrule/chondrules

An open question is whether micro-chondrules in Wild 2 share a common formation mechanism with chondrules in meteorites, or whether the only feature they share is the fact that they were once molten. The distribution of cations between phases in these objects may constrain their formation environment. In three micro-chondrules that have been studied in this way, cooling times and formation environments appear to have been similar to formation conditions of chondrules (Gainsforth *et al.* 2015). Another area of active investigation is whether determination of crystal site occupancy of cations through ALCHEMI may constrain formation conditions. **Many more analyses, including oxygen isotope studies, are needed to assess the range of formation conditions.**

Micro-CAIs
Based on textures, mineralogy and mineral chemistries, the Wild 2 fine-grained refractory particles appear to be most closely related to fine-grained inclusions which are commonly found in CV3 and more rarely in CO and CR chondrites and the primitive ungrouped carbonaceous chondrite Acfer 094.  Particle sizes of the Wild 2 refractory inclusions range from ~1 to 20 μm, the latter size representing the terminal particle from track 25 Inti.  These sizes likely represent lower bounds as fragments may have disaggregated during capture.  In general, refractory grains were found in about one-fifth of all tracks and in nearly 50% of all bulbous tracks.  Considering all studied large fragments from 19 tracks, a rough estimate indicates that ~2% of Wild 2 fragments are CAIs.   The Wild 2 CAI population appears to lack the most refractory CAI types such as Type A, hibonite-spinel and grossite-bearing varieties.  Nor have the highest temperature minerals – corundum $Al_2O_3$, hibonite $(Ca,Ce)(Al,Ti,Mg)_{12}O_{19}$, grossite $CaAl_4O_7$ – been found in Stardust tracks.   The high temperature mineral perovskite was observed as a rare phase in the terminal particle Inti from track 25 but has not been observed elsewhere.  At present, the reason for this segregation is unclear.  Resolving this uncertainty is a high priority for future research.  **Key will be to identify more micro-CAIs for study; high-resolution synchrotron X-ray microprobes may be particularly useful for this project.**

Cometary metals

Wild 2 contains large metal particles, with diverse chemical compositions.   One particle, Simeio, showed an unusual composition, with a very low Ni/Fe ratio and a trace element pattern most consistent with ureilitic metal (Humayun et al. 2015).  Metal grains have also been seen in numerous olivine grains and most spectacularly in the 40 μm chondrule, Gozensama, the first chondrule fragment identified in Wild 2 (Nakamura et al. 2008a).  Additional metal particles have been identified by synchrotron XRF and XANES analysis, but more particles will be needed to determine distributions for comparison with meteoritic metals.  Particle handling and sample preparation are particularly challenging for metal particles.  **Sample preparation techniques are currently in development for this purpose that do not contaminate samples with critical elements (Pt, W, Ga) (Burnett et al. 2016).**

**Objective 7: Determine origin of fines**

| Question and Objective | measurement | instrument | sample requirements | sample frequency |
|---|---|---|---|---|
| 7 How much interstellar matter is preserved in comets? *Determine origin of fines* | Mean and dispersion of elemental composition of fines | µXRF, SEM/EDX | keystone, craters in foils | common |
| | O-isotopic composition of fines | SIMS | dissected keystone, craters in foils | common |
| | Presolar grain concentration in fines | nanoSIMS | keystone, craters in foils | rare |
| | Isotopic anomalies, e.g., $\delta^{50}$Ti vs $\delta^{54}$Cr, $\delta^{50}$Ti vs $\delta^{46}$Ti, $\delta^{62}$Ni vs $\delta^{54}$Cr in fines | RIMS | shielded material (possibly behind terminal particles), craters in foils | rare |

**Summary of objectives**

- **Substantially improve statistics of oxygen-isotopic composition measurements of fines**
- **Determine bulk elemental and isotopic composition of fines for comparison with CI abundances**

**Technology advances**

- **Develop technique for laboratory hypervelocity capture experiments using fines, with gentle acceleration**

The fines (submicron components) are very important because they are a major fraction and perhaps the dominant fraction of the non-ice component mass of Wild 2. They will also take on increasing importance to future studies of Wild 2 samples because the collection contains nearly an unlimited number of grains smaller than a few microns. They have not been adequately studied or even adequately searched because they were not "low hanging fruit". They are difficult to study because they are small and they were commonly altered by capture heating and contact with molten silica. It is critical that work on these small components adequately distinguishes primary properties with those altered by capture. Most of the materials preserved as coarser grains have also been seen in fines, but it is possible that there were some fine components that did not survive capture. One possible way to make progress on the fines is to study the very smallest bulbous tracks (< 50 µm long) where thermal heating may have been significantly less than for the larger ones.

Because they contain relatively large phases, the largest Stardust particles are the easiest to extract and prepare, are the best-preserved, and contain relatively large phases, the largest particles have been the subject of the most intense investigation. However, there are indications that the fine-grained components of Wild 2 particles could be systematically different from the coarse-grained components. This may not be unexpected, if the coarse-grained components are "xenoliths" derived from the inner solar-system, and the fine-grained components are the relatively unprocessed solids that were present in the Kuiper Belt at the time of comet formation. The strongest evidence for such a systematic difference between coarse-grained and fine-grained material comes from oxygen isotopes, in which fine-grained material in one track shows a very large spread consistent with mixing with a very $^{16}$O-poor component like cosmic symplectite (Ogliore et al. 2015). **If confirmed through more analyses, this would lend strong evidence for a two-reservoir model to explain one of the key unsolved problems in planetary science:** *what is the origin of the pattern of oxygen isotopic composition of the Solar System, in the three-isotope oxygen diagram?*

The best opportunity for study of unaltered fine-grained material may be in a few rare cases in which it has been protected from damage during aerogel capture by proximity to a large, mechanically robust particle. So far two such cases, Febo and Andromeda, have been found, and in both cases, a field of fine-grained material appears on the downstream side of a large sulfide particle, although the pristinity of the fine-grained material related to Febo has been questioned (Ishii et al. 2015). Fig. 5 shows an SEM view of Andromeda with a large sulfide on the left and the FGM on the right. **Technique development is needed to identify such assemblages before embedding in epoxy or acrylic and ultramicrotomy, to avoid alteration, mobilization, and contamination of organics.** Although materials captured in aluminum foils are generally more severely altered than in aerogel, these materials may also be a resource for study of fine-grained materials. It is known that a fraction of submicron pre-solar grains survive capture in aluminum foils at ~6 km/sec capture speed. (Haas et al. 2016)

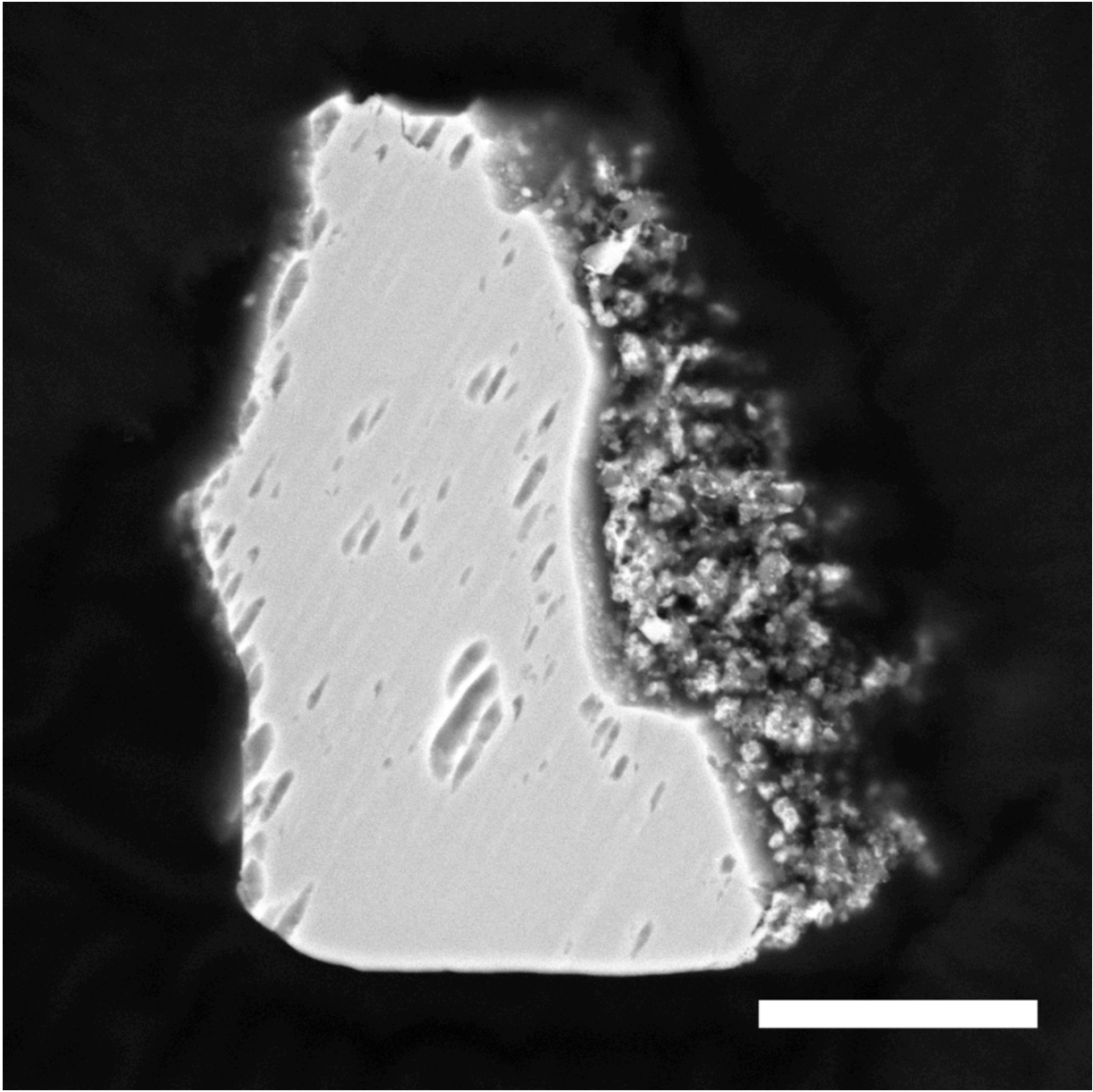

Figure 5: SEM image of the terminal particle named Andromeda. The large sulfide on the left may have partially shielded fine-grained materials during capture in aerogel, including crystalline and amorphous silicates and sulfides.

**Objective 8: Determine pre-accretionary space exposure, shock history, and ion irradiation histories**

| Question and Objective | measurement | instrument | sample requirements | sample frequency |
|---|---|---|---|---|
| **8 What is the history of cometary materials before comet formation?** *Determine pre-accretionary space exposure, shock history, ion irradiation, and thermal metamorphism* | Solar flare track concentration in silicates | TEM | not yet detected | rare |
| | Noble gas concentration, composition, and carrier | noble gas spectrometry | keystones | unknown |
| | Concentration of shocked phases | µXRD, analytical TEM, SXRD | keystones, u/m sections | common |
| | Concentration of products of secondary thermal processing | analytical TEM | u/m sections | rare |
| | Cosmogenic nuclides | AMS | | |

**Summary of objectives**

- **Substantially improve statistics in search for solar-flare tracks, and determine whether tracks survive aerogel capture.**
- **Substantially improve statistics in search for shocked phases**
- **Address reasons for current variability of noble gas concentrations and compositions in Wild 2 cometary materials and capture media, through expanded studies in multiple laboratories**
- **Determine whether the mineralogical criteria for identification of thermal metamorphism, which has been formulated principally for ordinary chondrites (OCs) and carbonaceous chondrites (CCs), can be applied to fine-grained materials in comets.**

**Technology advances**

- **Develop technique for laboratory hypervelocity capture experiments with gentle acceleration to determine whether aerogel capture can induce shock effects in minerals**

Space exposure

A determination of the space exposure time of the components of Wild 2 would be of great interest in the study of transport in the disk. **To date, no solar flare tracks have been found in any of the Wild 2 particles, suggesting that they were never exposed to solar flare**

**particles, but no systematic search has yet been done.** If tracks were not lost during capture, this suggests that all transport occurred when the disk retained gas that blocked energetic solar particles. Particles that were transported to the inner edge of the nebular disk might have been propelled from the innermost regions of the disk where they could have been exposed to high radiation environments, different oxidation states, and accretion of exotic components. An example of possible exotic effects close to the Sun might explain how nanophase osbornite TiN, Ti,V nitrides, and Platinum Group Elements nuggets became incorporated inside some of the Wild 2 CAIs. Osbornite condensation is predicted to only occur from gas with a C/O about twice the solar ratio.

Shock

Tomeoka *et al.* (2008), Jacob *et al.* (2009), Joswiak *et al.* (2012), and Gainsforth *et al.* (2015) reported shock effects in olivines and pyroxenes, which could be interpreted as evidence for mild shock. However, Stodolna et al. (2012b) showed that shock up to 1.5 GPa is produced by the combination of light gas gun, aerogel capture, and ultramicrotomy. Therefore it is not clear to what extent shock is present or absent in the comet without better characterizing the cause of the shock effects observed in the Stardust samples. A high priority is to confirm (or not) that some Wild 2 samples have undergone shock processes, because this has major implications for comet assembly or possibly cometary fragmentation as suggested by Farinella and Davis (1996) and Davis and Farinella (1997) . **This will require much more extensive TEM analyses of laboratory materials shot into aerogel as well as Wild 2 cometary samples, and possibly improvements in laboratory hypervelocity capture experiments.**

Noble gases

The noble gas concentration measured in a few Wild 2 particles is extremely high, higher than found in IDPs, but many Wild 2 particles are noble gas free (Marty *et al.* 2008, Palma *et al.* 2013). These high concentrations may be due to exposure of particles to solar energetic particles during the Sun's active phase, before incorporation into the comet. Strikingly high Ne concentrations are present in the track 41 particles, comparable to or exceeding those implanted into lunar fines and many IDPs by solar wind ions and suggesting gas acquisition by ion irradiation, in this case by a compositionally non-solar ion flux, for the track carrier grains (Marty et al., 2008). The noble gas patterns measured so far show no coherent pattern, so the noble gas carrier or carriers are unknown. To further complicate matters, some aerogel sections that are apparently track-free show the unexpected presence of noble gases, perhaps derived from nearby tracks (Palma *et al.* 2012).

The principal finding of these studies is the firm identification of the Q-Ne isotopic signature in Wild 2 coma grains, and the implication of this observation that Q-gases existed in the Solar System early enough in its history to be incorporated into comets. Apart from that, there are two outstanding issues to be addressed in future Stardust measurements. Noble gases measured in the subsurface block samples from cell C2044 are isotopically extremely variable and show no coherent pattern, nor are they associated with any detectable carrier grains.

Moreover the locations of these noble gas enclaves in C2044 are several millimeters removed from both the cell surface and the axis of track 41, well beyond the penetration range of solar wind ions, and there are no visible aerogel damage tracks connecting them to either the surface or the track. Even if some mechanism could be invoked to eject trackless and unobserved fragments from the track bulb wall to distant block aerogel positions, the wide compositional variability of the He and Ne in these samples argues against their origin from the isotopically Q-like track 41 impactor.  **Further analyses are needed to elucidate the origin of the variability.**

Thermal metamorphism

The low-Ca pyroxene/olivine ratio and their compositions are also sensitive indicators of the physico-chemical conditions of grain formation, permitting elucidation of cooling rate, degree of aqueous alteration, and thermal metamorphism.  Olivine and low-Ca pyroxene minor element compositions particularly provide evidence for thermal metamorphism having occurred (Frank et al. 2014). However, **an open question is whether the mineralogical criteria for this, which has been formulated principally for ordinary chondrites (OCs) and Carbonaceous Chondrites (CCs), can be applied to fine-grained materials in comets. Future TEM work on OC and CC matrices could settle this issue.**

**Objective 9:  Determine flux, size and velocity distribution, composition, and structure of local interstellar dust in the inner Solar System**

| Question and Objective | measurement | instrument | sample requirements | sample frequency |
|---|---|---|---|---|
| **9 What is the nature of local interstellar dust?  Determine flux, size and velocity distribution, composition and structure of interstellar dust in the inner Solar System** | Track density, elemental composition, min/pet and isotopic composition of IS candidates | optical microscopy, SEM/EDX, massively distributed search, STXM, µXRF/XRD, nanoSIMS, SIMS, RIMS | picokeystones, foils on stretchers, O isotope sample prep for tracks in development | ultrarare |

**Summary of objectives**
- **Substantially improve statistics of identified interstellar dust candidates in aerogel**
- **Substantially improve statistics of identified interstellar dust candidates in aluminum foils**
- **Determine mass, mineralogical and elemental composition of interstellar candidates in aerogel**

- **Determine mass, mineralogical and elemental composition of interstellar candidates in foils**
- **Measure oxygen isotopic abundances in ISD candidates in aerogel**
- **Measure oxygen isotopic abundances in ISD candidates in Al foils**

**Technology advances**
- **Development ultra-low risk technique for sample preparation for oxygen isotopic composition measurement of ISD candidates in aerogel**
- **Improve precision of O-isotopic composition measurements of residues in ISD craters in foils**
- **Develop capability of isotopic measurement of minor and trace elements in ISD candidates**

In addition to the goal of capturing and returning of cometary materials from the Jupiter-family comet Wild 2, Stardust had the goal of capturing and returning the first solid samples from the local interstellar medium.  To that end, a separate collector was exposed to the interstellar dust stream for ~200 days in two episodes, before the encounter with comet Wild 2.  This  is the smallest curated collection in the JSC curatorial facility by orders of magnitude.  Despite their size, these samples are extraordinarily valuable both because of their extreme rarity and because they are likely to be the first solid local interstellar materials ever returned to terrestrial laboratories for study.

A consortium (the Interstellar Preliminary Examination, ISPE) (Westphal et al., 2014b) was formed to identify and analyze candidate interstellar dust impacts in the aerogel tiles (Westphal et al., 2014a, Frank et al. 2013, Bechtel et al. 2013, Butterworth et al. 2014, Brenker et al., 2013, Simionovici et al. 2014, Flynn et al. 2014, Gainsforth et al. 2014, Postberg et al. 2014, Sterken et al. 2014) and aluminum foils (Stroud et al. 2014).  This project resulted in the discovery of seven particles of likely interstellar origin, and also provided an upper limit on the flux of interstellar dust in the Solar System that is substantially lower than was expected based on the observations of interstellar dust by the Ulysses and Galileo spacecraft (Westphal et al. 2014c).   Since the end of the ISPE, scanning and identification of tracks in aerogel and craters in foils has slowed, principally because of limited resources at the curation facility at Johnson Space Center.  Nevertheless, impact identification is continuing, through the Stardust@home distributed search (Westphal et al. 2014a) and, recently, through a parallel effort called foils@home to identify impacts in foils (Stroud et al. 2016).  A complete identification of all impacts will require several more years.   Despite the effort, the project is clearly scientifically compelling:  this collection has the potential to contain the first interstellar dust particles ever identified and analyzed in terrestrial laboratories.  (It is important to draw a distinction between *interstellar dust* -- tiny particles currently residing in the local interstellar medium, probably no more than a few hundred Ma old -- and *circumstellar dust* -- condensates from circumstellar outflows, which are probably a small component of the interstellar dust.  The only samples of circumstellar dust unambiguously recognized so far are the so-called presolar grains -- ancient circumstellar grains that pre-date the Solar System and became incorporated into planetary materials.)

The evidence for interstellar origin for the seven candidates identified so far is strong (Westphal et al. 2014c) but circumstantial. Recently Silsbee and Draine showed that two of the largest particles, Orion and Hylabrook, require special optical properties to be consistent with their apparently low capture speed in aerogel (Silsbee and Draine 2016). The mean isotopic composition of oxygen in the interstellar medium differs significantly from that of the solar system. **A robust test of interstellar origin requires a measurement of oxygen isotopic composition of these candidates and others that are identified in the future with sufficient precision and accuracy that they can be distinguished from particles of solar-system origin**. But we should note that IS grains could be isotopically solar within measurement precision. Measurements can be carried out now in particles captured in the aluminum foils of the collector, but the error bars are not sufficiently small to determine their origins. For the particles in aerogel, the challenge is different: the measurement requires that the particles be exposed, but **no reliable technique exists yet for preparing ~1µm particles captured in aerogel for isotopic analysis in a SIMS instrument** (Westphal et al. 2016b).

The CHILI instrument (Stephan et al, 2016) is ideally suited for analyses of ISD particles, because of its extraordinarily efficient use of the samples. Work is ongoing to develop sample preparation techniques for ISD analysis with CHILI. The Al-Mg chronometer is a likely target for technique development.

**SUMMARY**

The comet Wild 2 samples returned by Stardust have provided information on the early Solar System that could not have been known without actual proven comet samples in the laboratory. The low abundance of isotopically anomalous grains, and the presence of CAIs and abundant chondrule fragments has profoundly influenced the understanding of the formation of comets and icy planetesimals and has provided evidence of large scale movement of common asteroid building materials across the Solar System. While the program has greatly exceeded common expectations for the low cost Discovery 4 mission, there is much more that needs to be done. The next comet sample return is well off in the future and the Wild 2 material will remain a unique resource for years to come. The use of refined handling and analytical methods, researchers that are well experienced with dealing with these small complex samples, the accumulated knowledge needed to recognize alteration from high-speed capture and the accumulated insight into important issues that can be studied with these samples: all provide an excellent base from which to launch future studies. To optimize the science return from the Stardust mission, these studies should be coordinated in ways that maximize the science return from each track and crater because each particle contains information rich-materials with very important stories to tell. The finding of CAIs, $^{16}$O rich phases, an excellently preserved isotopically anomalous pre-solar SiC grain, and good candidates for dating using short-lived isotopes were all done on minor components of capture tracks that could easily have been overlooked, lost, or destroyed in the samples studied during the early years of Wild 2 studies. A major lesson from the comparison of the organized consortium-based approach applied to the Interstellar side of the collector with the more traditional approach to analyses of the comet side is that coordinated approaches with planned focus are the most productive and the best use of these precious samples. Well-coordinated future studies should be able to answer some of the most burning questions on Wild 2 materials such as determining if the Wild-2 chondrule-like and CAI-like components are derived from the same source regions as their chondrite counterparts or if they were just made by similar nebular high temperature processes and may have come from different locales plausibly at different times.

A major point not discussed in this paper is the question of how representative Wild 2 is of Jupiter Family comets. We know that volatiles vary from comet to comet but what about the rocky materials that actually dominate the mass of most comets? If all rocky materials were transported to the comet accretion region, then they might have similar mixes of materials. If comets accreted over an extended period of time, there might be comets that contain different mixes of materials transported from different regions or different times. One way to address this is to use Wild 2 to identify IDPs or even meteorites that come from other comets. This is a bit of a bootstrap but it might be done simply by comparing the compositional dispersion of accreted materials. One interpretation of the broad range of materials in Wild 2 is due to its formation by scavenging material from dispersed environments while chondrite classes are distinctive because they contain large amounts of regionally produced components.

Acknowledgments

This paper is the product of a workshop that was held near the UC Berkeley campus on July 26-27, 2015, just prior to the 78th annual meeting of the Meteoritical Society. The workshop was supported by NASA and sponsored by the Curation and Planning Team for Extraterrestrial Materials (CAPTEM), a NASA advisory committee

# Glossary of Experimental Techniques

| Acronym | Technique | Incident particles | Signal | Science Goal |
|---|---|---|---|---|
| **ALCHEMI** | Atom Location by CHanneling Enhanced MIcroanalysis | Electrons | Electrons | Site occupancies |
| **AMS** | Accelerator Mass Spectrometry | | | Isotopic composition |
| **APT** | Atom Probe Tomography | photons | Ions | Nano morphology, element abundances and isotopic ratios |
| **EDX or EDS** | Energy-dispersive X-ray Spectroscopy | Electrons, ions, or X-ray photons | X-ray photons | Sample composition (similar to XRF) |
| **EELS** | Electron Energy-Loss Spectroscopy | Electrons | Electrons | Chemical environment of a specific element (similar to XANES) |
| **EPMA** | Electron Probe Microanalysis | Electrons | Electrons + X-ray photons | Sub-micron morphology and composition |
| **FIB** | Focussed Ion Beam Milling | Ions | Electrons and ions | Sample Preparation |
| **FTIR** | Fourier Transform Infrared Spectroscopy | Infrared photons | Infrared photons | Bonds |
| **HIM** | Helium Ion Microscopy | Ions | Electrons and ions | Particle structure |
| **HRTEM** | High-resolution Transmission Electron Microscopy | Electrons | Electrons | Crystal structure and atomic scale defects |
| **L$^2$MS** | Two-Step Laser Desorption Mass Spectrometry | photons | Ions | Trace element abundances and isotopic ratios |
| **LA-ICP-MS** | Laser Ablation — Inductively Coupled Plasma Mass Spectroscopy | photons | Ions | Trace element abundances and isotopic ratios |
| **LGCMS** | Liquid Gas Chromatograph Mass Spectroscopy | | | Composition of organics |
| **nanoFTIR** | nanometer resolution FTIR | Infrared photons | Infrared photons | Bonds |
| **nanoSIMS** | nanometer resolution Secondary Ion Mass Spectrometry | Ions | Ions | Isotopic ratios of elements |
| **noble gas analysis** | Noble Gas Mass Spectrometry | | | Noble gas concentration and isotopic ratios |

| pSTXM | Ptychography/STXM | X-ray photons | X-ray photons | Nanoscale morphology and chemistry (similar to XANES) |
| --- | --- | --- | --- | --- |
| RIMS | Resonance Ionization Mass Spectrometry | Ions or photons | Ions | Isotopic ratios of elements |
| SEM/EDX | Scanning Electron Microscope with Energy Dispersive X-ray Spectroscopy | Electrons | Electrons + X-ray photons | Sub-micron morphology and composition |
| SIMS | Secondary Ion Mass Spectrometry | Ions | Ions | Isotopic ratios of elements |
| (S)TEM/EDS | (Scanning) Transmission Electron Microscopy with Energy Dispersive X-ray Spectroscopy | Electrons | Electrons + X-ray photons | Nanoscale morphology and composition |
| SXRD | Synchrotron-based X-ray Diffraction | X-ray photons | X-ray photons | Crystal structure |
| SXRF | Synchrotron-based X-ray Fluorescence Spectroscopy | X-ray photons | X-ray photons | Sample composition (similar to EDS) |
| TOF-SIMS | Time-of-Flight Secondary Ion Mass Spectrometry | Ions | Ions | Elemental and isotopic ratios, organics |
| XANES | X-ray Absorption Near-Edge Spectroscopy | X-ray photons | X-ray photons | Chemical environment of a specific element (similar to EELS) |